\documentclass[conf]{new-aiaa}
\usepackage[utf8]{inputenc}

\usepackage{graphicx}
\usepackage{float}
\usepackage{amsmath}
\usepackage[version=4]{mhchem}
\usepackage{siunitx}
\usepackage{longtable,tabularx}
\usepackage{bm}
\usepackage{siunitx}
\usepackage{caption}
\usepackage{subcaption}

\title{A Machine Learning Approach to Classify Kinematics and Vortex Wake Modes of Oscillating Foils}

\author{Bernardo Luiz R. Ribeiro\footnote{PhD student, Engineering Physics, AIAA Member.}}
\author{Jennifer A. Franck\footnote{Assistant Professor, Engineering Physics, AIAA Senior Member.}}
\affil{University of Wisconsin-Madison, Madison, WI, 53706}

\begin{document}

\maketitle

\textbf{Machine learning techniques have received attention in fluid dynamics in terms of predicting, clustering and classifying complex flow physics. One application has been the classification or clustering of various wake structures that emanate from bluff bodies such as cylinders or flapping foils, creating a rich diversity of vortex formations specific to flow conditions, geometry, and/or kinematics of the body. When utilizing oscillating foils to harvest energy from tidal or river flows, it is critical to understand the intricate and nonlinear relationship between flapping kinematics and the downstream vortex wake structure for optimal siting and operation of arrays. This paper develops a classification model to obtain groups of kinematics that contain similar wake patterns within the energy harvesting regime. Data is obtained through simulations of 27 unique oscillating foil kinematics for a total of 13,650 samples of the wake vorticity field.  Within these samples three groups are visually labeled based on the relative angle of attack. A machine learning approach combining a convolutional neural network (CNN) with long short-term memory (LSTM) units is utilized to automatically classify the wakes into the three groups. The average accuracy on five test data subsets is 80\% when the three visually labeled groups are used for classification. After analyzing the test subset with lowest accuracy, an update on the group division boundaries is proposed. With this update, the algorithm achieves an average accuracy of 90\%, demonstrating that the three groups are able to discern distinct wake structures within a range of energy harvesting kinematics.}

\section{Introduction} \label{intro}

This paper utilizes a machine learning approach to classify vortex wake structures behind an oscillating foil. Understanding the vortex formation and resulting wake structure reveals information about the upstream disturbance, which could potentially be linked to the underlying flow conditions and/or kinematics of the oscillating foil. Although commonly used in propulsive applications oscillating foils can also act as energy harvesters, extracting energy from the incoming flow in a similar manner as a rotational turbine \cite{Young2014, Xiao2014}. Due to the coherent alternating signed vortex wake, there is strong potential for oscillating foils to work cooperatively within tightly packed array configurations by taking advantage of the structured wake to enhance performance. In order to create the control laws to optimize performance in array configurations, one needs to fully understand and model the wake structure as a function of flow conditions and foil kinematics. 

Wakes of bluff bodies have been most commonly investigated for cylindrical geometries, such as the canonical work of Williamson and Roshko \cite{williamson1988} who describe vortex wakes with a `\textit{mS + nP}' notation, where \textit{m} is the number of single vortices (\textit{S}) shed per cycle, and \textit{n} is the number of clockwise/counter-clockwise vortex pairs (\textit{P}). This notation has been propagated to oscillating foils with some success \cite{schnipper2009}, and others have thoroughly investigated wake structures behind both pitching \cite{koochesfahani2012} and plunging motions \cite{lai1999} primarily for propulsive applications. When used for energy harvesting, the foil kinematics are characterized by a lower non-dimensional frequency and higher pitch/heave amplitudes compared to oscillating foil propulsion. The high amplitudes result in rich wakes of multiple alternative sign vortices shed each foil stroke that are often chaotic and not easily identified by the `\textit{mS + nP}' notation \cite{RibeiroFranck2020}.

%However, there is not much work on foils in energy harvesting mode in terms of identifying wake patterns. The foil kinematics in this mode are characterized by a lower non-dimensional frequency and higher pitch-heave amplitudes compared to oscillating foil propulsion. This results in rich wakes of multiple alternative sign vortices shed each foil stroke that are not easily identified by the canonical wake characterization \cite{williamson1988}. Oscillating foils in energy harvesting mode are especially useful if placed in array configuration as the wake between foils can enhance the energy harvesting on downstream foils depending on the array setup \cite{platzer2009, Ashraf2011, broering2012b, kinseytandem2012, RibeiroFranck2021}. A major challenge in this field is to associate foil kinematics with the wake generated since each foil parameter changes the vortex formation in a different way \cite{RibeiroFranck2020}. Wake identification strategies, and correlation with the foil kinematics, would enable more sophisticated control schemes in which array configurations can work efficiently.

In terms of characterizing wakes structures of bluff bodies, various machine learning techniques can be utilized. 
%In terms of identification strategies, machine learning has shown to be an efficient tool in addressing many challenges in fluid mechanics \cite{brunton2020}, including those related to control and wake characterization. Machine learning is defined as the field of study that gives machines the ability to learn without using an explicit program code \cite{agostini2020}. This field is divided in three categories, namely unsupervised, semi-supervised and supervised learning \cite{brunton2020}. Unsupervised learning does not use labeled data and consists of learning unknown patterns, which are commonly the target of clustering algorithms \cite{calvet2020, zhang2019} or self-organizing maps \cite{wu2019}. The semi-supervised learning involves the technique called `reinforcement learning', which was extensively investigated in understanding fish schooling formations \cite{Verma2018, gazzola2014}, swimming behavior \cite{jiao2021} and aircraft formation flight \cite{ransquin2021}. Regarding supervised learning, one common application is prediction of flow fields through neural networks and compare against numerical simulations \cite{ti2020}.
Particularly, convolutional neural networks (CNN) have received much interest due to the ability to process data from images for pattern recognition and prediction \cite{brunton2020}. For instance, this type of neural network helped develop a model to analyze unsteady flow over a circular cylinder at various Reynolds numbers and obtained accurate predictions \cite{lee2019}. Similarly, Bhatnagar et al. predicted flow fields around various airfoil shapes \cite{bhatnagar2019}. A CNN was also utilized to develop a vortex identification technique that could potentially replace existing identification methods such as Q-criterion \cite{Hunt1988} or $\lambda_2$ criterion \cite{jeong1995}, which typically require user-input for appropriate thresholds \cite{deng2019}. Calvet et al. developed a convolutional autoencoder (CAE) that extracted the most pertinent features of an oscillating foil in propulsion mode using it to automatically cluster wake patterns from various foil kinematics \cite{calvet2020}. Using a semantic segmentation approach within a CNN, Kashir et al. \cite{kashir2021} successfully applied feature extraction to a lid-driven cavity flow. CNNs have also been used to predict drag and lift coefficients on a flat plate and on two side-by-side cylinders \cite{morimoto2021}.

When analyzing unsteady flows, the time evolution of structures is relevant in capturing the subtle differences between present and past timesteps. Recurrent neural networks and especially their improved version, the long short-term memory (LSTM) method \cite{hochreiter1997}, have been shown to predict flow structures based on past instances. The internal memory of a LSTM unit holds information on the recent context in an input sequence, and not simply from the current input to the neural network \cite{han2019}. %Since the LSTM method is able to learn from data that passed through the network, researchers have used this method to predict unsteady flow structures. 
Using flow information from the past five timesteps, Nakamura et al. \cite{nakamura2021} predicted turbulent structures in a channel flow using a convolutional autoencoder combined with LSTM. LSTM was also used to develop a novel dynamic wind farm wake model that predicts the main features of unsteady wind turbine wakes almost as well as high-fidelity wake models \cite{zhang2020}. Using convolutional layers and LSTM units, Han et al. \cite{han2019} predicted unsteady flow fields behind bluff bodies such as a cylinder and a foil.

Although combining CNN and LSTM have been successful in predicting unsteady flows, the purpose of this paper is to combine these two techniques towards classification of flow fields. Classification models have been implemented in oscillating foil propulsion \cite{colvert2018, pollard2021} and behind cylinders \cite{li2020}, which have shown to be successful on classifying wakes from point-measurements or from the foil's kinematics. In contrast, the goal of this paper is to classify vortex wake structures solely based on images, or vorticity flow fields, and correlate the classes with oscillating foil kinematics for energy harvesting. Thus, by understanding how the kinematics are associated with wake structure, predictive models of array configurations, kinematics, and performance can be constructed. Using two-dimensional direct numerical simulations (DNS), Ribeiro et al. \cite{RibeiroFranck2021} defined three modes for vortex wakes by measuring the strength of the first vortex shed from the suction side of the oscillating foil each half-stroke, but did not consider the entire wake structure. % However, with these thresholds to train the data, this paper automatically separates 13,650 samples from 27 oscillating foil kinematics into three modes, and uses these results to re-evaluate the boundaries between modes. 
%Although the analysis of the first vortex formation and shedding is relevant for energy harvesting in foil-arrays, its connection to wake patterns is still missing.

This paper develops a classification model based on convolutional layers and LSTM units, which is applied on wake structures captured by vorticity magnitude among a wide range of 27 foil kinematics for energy harvesting. Using visually labeled groups from Ribeiro et al. \cite{RibeiroFranck2021}, the classification model assesses if these groups are correctly divided or need adjustment based on the prediction accuracy of the model.

%With the possibilities that a classification model, CNN and LSTM provide in terms of identifying wake patterns and analyzing time evolution of flow fields, the application of those tools on oscillating foils in energy harvesting is proposed in this paper with the goal on associating foil kinematics with wake patterns, which will be useful for predicting energy harvesting efficiency in multiple-foil arrays. This paper focus on developing a classification model based on convolutional layers that extract spatial features of images containing fields of flow vorticity magnitudes and the application of LSTM units to analyze the time evolution of these spatial features among a wide range of foil kinematics in energy harvesting regime. Using predefined classes based on previous work by Ribeiro et al. \cite{RibeiroFranck2021}, the classification model assesses if these prelabeled groups are correctly divided or need adjustment considering model's accuracy. Two-dimensional direct numerical simulations (DNS) are used for this purpose.

\section{Computational Methods}

\subsection{Foil Parameters}

In order to generate a range of energy harvesting vortex wake structures 27 unique kinematics are prescribed to a $10\%$ elliptical foil through numerical simulations. The foil motion is sinusoidal in pitch and heave, utilizing a range of frequencies and amplitudes previously established as effective at energy harvesting \cite{kindum2008, RibeiroFranck2020}. As opposed to an airfoil geometry, a thin elliptical foil of chord length $c$ is utilized as foil shape has been shown to have only minor effects on efficiency \cite{Kim2017} and the fore-aft symmetry is desirable for harvesting energy from tidal flows. The kinematics are described in lab-fixed coordinates as

\begin{equation}
h(t)=h_{o}\cos(2\pi f t) 
\label{eq:heave}
\end{equation}

\noindent and 

\begin{equation}
\theta(t)=\theta_{o}\cos(2\pi f t + \pi/2),
\label{eq:pitch}
\end{equation}

\noindent where $h(t)$ and $\theta(t)$ are the prescribed heave and pitch motions, respectively, with pitch about the center-chord. Heave and pitch are changing simultaneously during foil motion, which creates a time-varying relative angle of attack with respect to the freestream flow given by

\begin{equation}
\alpha_{rel}(t) = \tan^{-1} (-\dot{h}(t)/U_{\infty}) + \theta(t),
\label{eq:alpharel}
\end{equation} 

\noindent with $\dot{h}(t)$ representing the time derivative of the heave motion and $\alpha_{rel}(t)$ is in radians. A characteristic relative angle of attack is evaluated when the foil is at maximum $\theta$ ($\theta_o$) and maximum heave velocity, which occurs at one quarter of the cycle period $T$, or

\begin{equation}
\alpha_{T/4} =\alpha_{rel}(t = 0.25T).
\label{eq:at4}
\end{equation}

\begin{figure}[htbp]
\centering
\includegraphics[width=0.27\textwidth]{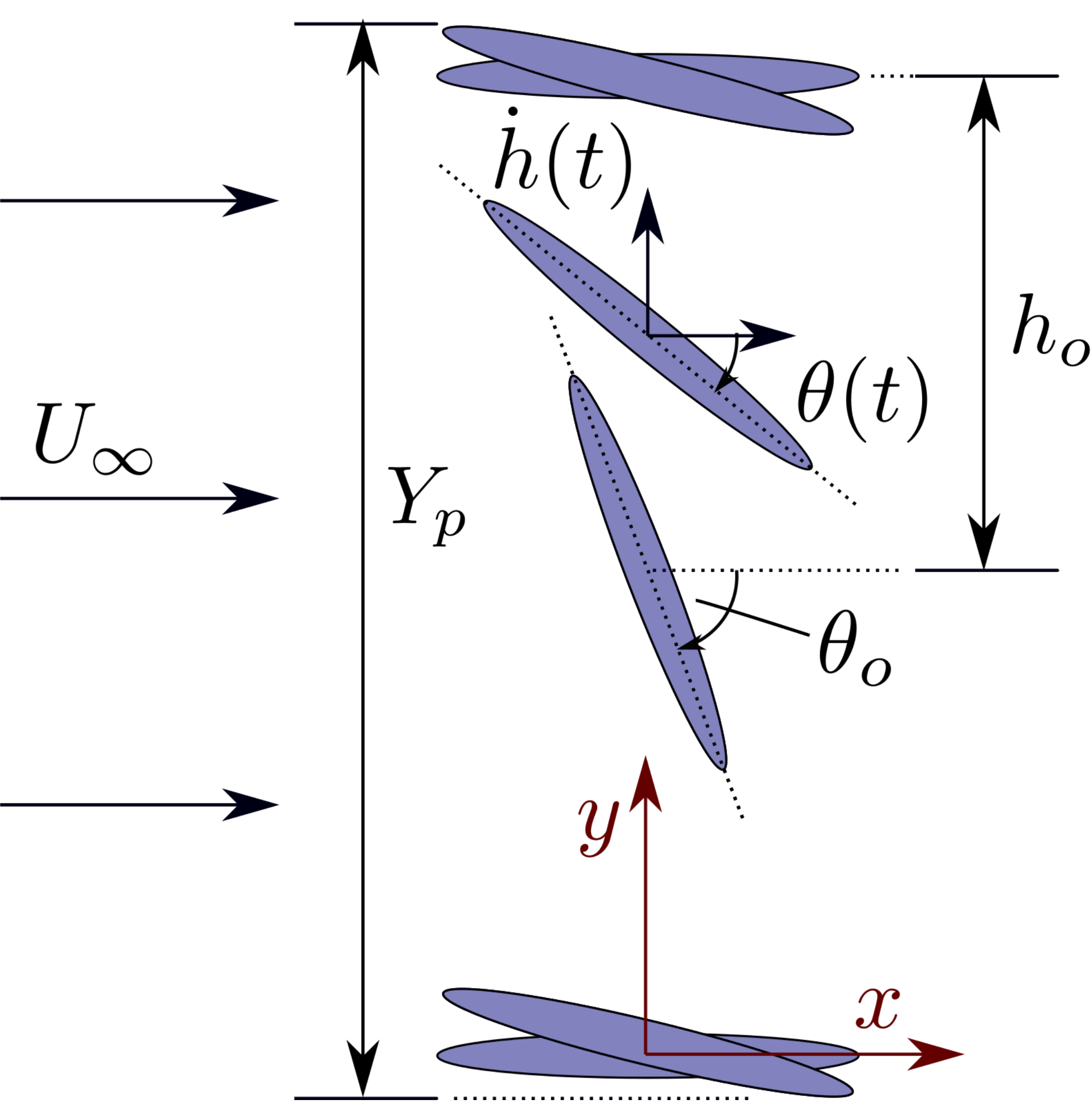}
\caption{Foil kinematics: pitch amplitude ($\theta_o$), heave amplitude ($h_o$) and foil's swept area ($Y_p$) shown.}
\label{f:singlefoil}
\end{figure}

The foil kinematics are thus defined by the three non-dimensional parameters (pitch amplitude, $\theta_o$, heave amplitude, $h_o$, and reduced frequency, $fc/U_\infty$) as shown in Figure \ref{f:singlefoil}. All quantities reported are non-dimensionalized by the freestream velocity, $U_\infty$, and $c$. 

To evaluate performance the energy harvesting efficiency is defined as

\begin{equation}
\eta = \frac{\bar{P}}{\frac{1}{2}\rho U_{\infty}^3 Y_p},
\label{eq:eta}
\end{equation}

\noindent which is the ratio of the average power extracted throughout a single stroke, $\bar{P}$, compared to the power available in the freestream velocity throughout the swept area $Y_p$. Another metric of performance is the power coefficient,

\begin{equation}
    C_p = \frac{\bar{P}}{\frac{1}{2}\rho U_{\infty}^3 c},
\end{equation}

\noindent whose denominator scales with the size of the foil, or the size of the energy harvesting device, rather than the intercepted flow area. Note that for traditional rotating turbines these two metrics are identical, but the variation in heave amplitude with fixed span and chord yields two diverse performance metrics for oscillating foils. The power extracted is defined as

\begin{equation}
P(t)=F_y \dot{h} + M_z \dot{\theta},
\label{eq:power}
\end{equation}

\noindent where $F_y$ and $M_z$ are the vertical force and spanwise moment on the foil respectively. To remove small cycle-to-cycle variations the efficiency and power coefficient are phase-averaged over the last three cycles of simulation.

This paper considers a wide range of kinematics all within the energy harvesting regime through variation in heave amplitude, pitch amplitude, and reduced frequency, all of which directly influence $\alpha_{T/4}$. Table \ref{t:kin} outlines the 27 kinematics, with three sets of different kinematics chosen specifically with the same $\alpha_{T/4}$.

\begin{table}[htbp]
\setlength{\tabcolsep}{8pt}
\centering
\begin{tabular}{cc|cc}
\hline
\textbf{Kinematics} & \textbf{$\alpha_{T/4}$} & \textbf{Kinematics} & \textbf{$\alpha_{T/4}$} \\ \hline
$fc/U_\infty = 0.15 ; h_o/c = 1.25 ; \theta_o = 55^{\circ}$ & 0.09 & $fc/U_\infty = 0.12 ; h_o/c = 1.25 ; \theta_o = 65^{\circ}$ & 0.38 \\
$fc/U_\infty = 0.12 ; h_o/c = 1.50 ; \theta_o = 55^{\circ}$ & 0.11 & $fc/U_\infty = 0.10 ; h_o/c = 1.25 ; \theta_o = 60^{\circ}$ & 0.38 \\
$fc/U_\infty = 0.15 ; h_o/c = 1.00 ; \theta_o = 50^{\circ}$ & 0.12 & $fc/U_\infty = 0.15 ; h_o/c = 1.25 ; \theta_o = 75^{\circ}$ & 0.44 \\
$fc/U_\infty = 0.12 ; h_o/c = 1.00 ; \theta_o = 45^{\circ}$ & 0.14 & $fc/U_\infty = 0.15 ; h_o/c = 1.00 ; \theta_o = 70^{\circ}$ & 0.47 \\
$fc/U_\infty = 0.10 ; h_o/c = 1.00 ; \theta_o = 40^{\circ}$ & 0.14 & $fc/U_\infty = 0.12 ; h_o/c = 1.00 ; \theta_o = 65^{\circ}$ & 0.49 \\
$fc/U_\infty = 0.12 ; h_o/c = 0.75 ; \theta_o = 40^{\circ}$ & 0.18 & $fc/U_\infty = 0.12 ; h_o/c = 0.75 ; \theta_o = 60^{\circ}$ & 0.53 \\
$fc/U_\infty = 0.15 ; h_o/c = 1.00 ; \theta_o = 55^{\circ}$ & 0.20 & $fc/U_\infty = 0.15 ; h_o/c = 1.00 ; \theta_o = 75^{\circ}$ & 0.55 \\
$fc/U_\infty = 0.10 ; h_o/c = 1.00 ; \theta_o = 45^{\circ}$ & 0.22 & $fc/U_\infty = 0.12 ; h_o/c = 1.25 ; \theta_o = 75^{\circ}$ & 0.55 \\
$fc/U_\infty = 0.12 ; h_o/c = 1.00 ; \theta_o = 50^{\circ}$ & 0.23 & $fc/U_\infty = 0.10 ; h_o/c = 1.00 ; \theta_o = 65^{\circ}$ & 0.57 \\
$fc/U_\infty = 0.10 ; h_o/c = 0.75 ; \theta_o = 40^{\circ}$ & 0.26 & $fc/U_\infty = 0.12 ; h_o/c = 0.75 ; \theta_o = 65^{\circ}$ & 0.62 \\
$fc/U_\infty = 0.15 ; h_o/c = 1.00 ; \theta_o = 60^{\circ}$ & 0.29 & $fc/U_\infty = 0.12 ; h_o/c = 1.00 ; \theta_o = 75^{\circ}$ & 0.66 \\
$fc/U_\infty = 0.12 ; h_o/c = 1.00 ; \theta_o = 55^{\circ}$ & 0.31 & $fc/U_\infty = 0.10 ; h_o/c = 1.00 ; \theta_o = 75^{\circ}$ & 0.75 \\
$fc/U_\infty = 0.10 ; h_o/c = 0.75 ; \theta_o = 45^{\circ}$ & 0.34 & $fc/U_\infty = 0.10 ; h_o/c = 0.50 ; \theta_o = 65^{\circ}$ & 0.83 \\
$fc/U_\infty = 0.12 ; h_o/c = 0.75 ; \theta_o = 50^{\circ}$ & 0.36 \\
\hline \\
\end{tabular}
\caption{Summary of all kinematics with their computed $\alpha_{T/4}$ values.}
\label{t:kin}
\end{table}

\subsection{Flow Solver and Mesh}

\begin{figure}[H]
\centering
\includegraphics[width=0.38\textwidth]{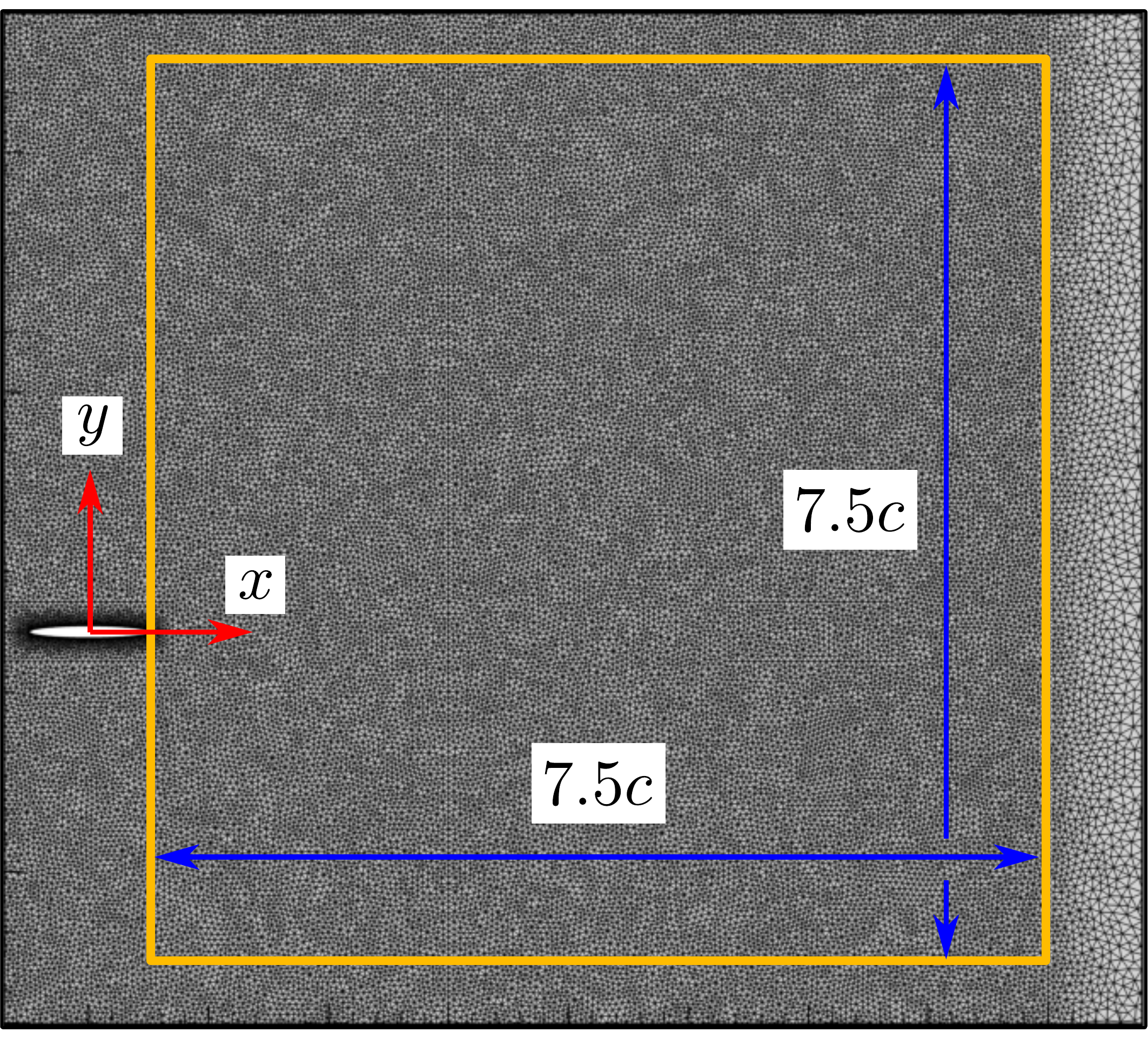}
\caption{Computational  domain  zoomed in on the foil mesh and the window used to extract the vorticity flow field.}
\label{f:mesh}
\end{figure}

The computations utilize an incompressible direct numerical simulation (DNS) at Reynolds number $Re=1000$ and are performed using a second-order accurate finite volume, pressure-implicit split-operator (PISO) method \cite{issa1986} implemented in $\textit{OpenFOAM}$ \cite{weller1998}. A 2D unstructured mesh is generated using Gmsh \cite{gmsh} with a subset of the mesh shown in Figure \ref{f:mesh}. The motion of the foil is generated through the boundary condition of a dynamic mesh solver, which updates the position of all nodes in the domain at every timestep. The total domain size is $106$ chord lengths ($106c$) in the horizontal direction with $100c$ in the vertical direction. The foil is located $50c$ upstream in the $x$ direction and is centered at foil's bottom stroke.

The mesh has approximately $1.64 \times 10^{5}$ points following a mesh refinement study found in full detail in Ribeiro et al. \cite{RibeiroFranck2021}. The boundary conditions entail a no-slip condition at the foil surface with zero pressure gradient, inlet flow on the left boundary, and outlet flow on the top, bottom and right boundaries. Simulations are run for a total of eight oscillation cycles, and are typically at steady state after three cycles.

\subsection{Machine Learning Classification Model} \label{mlmethods}

\subsubsection{Data Pre-Processing}

The input to the classification model are matrices of 2D vorticity magnitude of the wake behind the foil. The data is extracted from the simulations within a $7.5c$ by $7.5c$ window behind the foil as shown in Figure \ref{f:mesh} and interpolated to a cartesian grid of $128$ by $128$ points. The window size is selected such that it contains all vortices shed from the foil in all kinematics. For each set of kinematics three oscillation cycles within the steady state regime are used as input with data sampled every $tU_\infty/c=0.1$. 

With LSTM units being used in the classification model, tests are performed to find the ideal time sampling rate for the input sequence. It is found that $5$ time samples provide better overall accuracy than larger input sequences. To avoid overfitting, a data augmentation technique is implemented, which approximately duplicated the number of samples from $6827$ to $13,650$. This technique involves taking the $5$-sample input sequence not only at consecutive $0.1 tU_\infty/c$ units but also with $0.1  tU_\infty/c$ being skipped between samples, i.e. $0.1, 0.3, 0.5, 0.7, 0.9  tU_\infty/c$.

%Besides data augmentation, a $5$-fold stratified cross-validation is implemented in the proposed classification model. As noted by Fukami et al. \cite{fukami2020} and Brunton and Kutz \cite{bruntonbook}, this process is critically important to avoid overfitting. The data, comprised of matrices of the vorticity fields and their respective class labels, are then divided in five subsets where one subset is separated for testing while the others are used for training. The test subset is finally permuted until five combinations or `folds' of test/training data are obtained. The stratified component in the cross-validation process assures representative information for all three visually labeled classes in each test subset \cite{keras}.

\subsubsection{Classification Model Architecture} \label{s:modelarch}

\begin{figure}[H]
\centering
\includegraphics[width=1\textwidth]{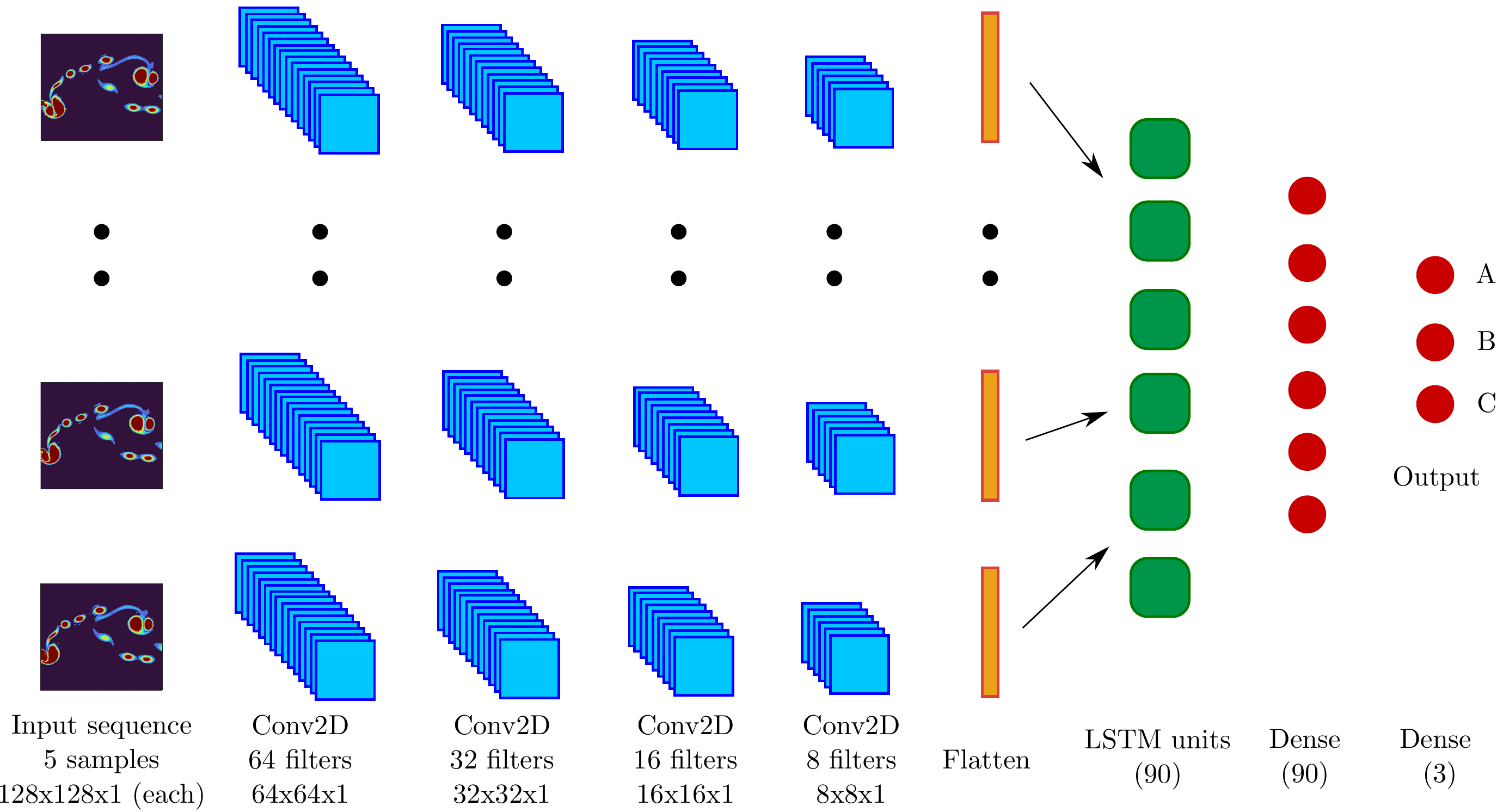}
\caption{Classification model architecture.}
\label{f:model}
\end{figure}

The classification model architecture is outlined in Figure \ref{f:model} and is implemented using the Python based libraries \textit{TensorFlow} \cite{tensorflow} and \textit{Keras} \cite{keras}. With an input sequence of $5$ samples, 2D convolutional layers (\textit{Conv2D}) are applied on each sample to extract the most significant features of the vortex wake. These key features are detected with filters, which create features maps through a convolutional operation on the preceding layer \cite{calvet2020}. Each convolutional layer uses a linear activation function and has multiple filters, which have a kernel size of $3\times 3$ and that decrease in number, thus reducing the matrix dimensions while simultaneously keeping the most pertinent features. The number of feature maps define the depth of each convolutional layer and within each layer, a downsample operation is performed with a $2\times 2$ stride, resulting in a reduction of factor $2$ in each matrix dimension while the depth remains constant.

Each input sequence passes through four convolutional layers, decreasing the dimension of each sample from $128\times 128 \times 1$ to a $8 \times 8$ feature map with $8$ channels. After reaching the final size, each sample is flattened into a 1D vector with $512$ elements and then inserted into $90$ LSTM units, with the goal of analyzing the time evolution of the most pertinent features captured by the convolutional layers. Finally, $90$ neurons in a dense layer are connected to the LSTM units and then a final dense layer of three neurons outputs each sample to the corresponding class `A', `B' or `C'. Both dense layers use a sigmoid activation function as it is commonly used in many machine learning models to normalize the output from previous layer into a $0-1$ range. %And therefore, values within this range can be interpreted as a probability function. %The number of filters, LSTM units and number of neurons in the dense layer were appropriately tuned.

The following hyperparameters in the classification model were tuned: Number of filters in the convolutional layers, LSTM units, and number of neurons in the first dense layer. The sequences of $(64,32,16,8)$, $(32,16,8,4)$, $(128,64,32,16)$ filters for the convolutional layers were tested and $(64,32,16,8)$ obtained best performance. The LSTM units and number of neurons in the first dense layer were tuned using a range from $20-100$ units and neurons. It was found that $90$ units and $90$ neurons provided a higher model accuracy. The \textit{Adam} optimization algorithm \cite{adam} is implemented to iteratively compute the parameter values for the entire classification model and to decrease the loss function, which uses categorical crossentropy \cite{keras} for the multi-class model with three outputs.

To avoid overfitting, a $5$-fold stratified cross-validation is implemented in the proposed classification model. As noted by Fukami et al. \cite{fukami2020} and Brunton and Kutz \cite{bruntonbook}, this process is critically important. The data, comprised of matrices of the vorticity fields and their respective class labels, are then divided in five subsets where one subset is separated for testing while the others are used for training. The test subset is finally permuted until five combinations or `folds' of test/training data are obtained. The stratified component in the cross-validation process assures representative information for all three visually labeled classes in each test subset \cite{keras}. Besides stratified cross-validation and data augmentation, the early stopping technique and a dropout layer with rate equal to $0.1$ between the dense layers are also used to avoid overfitting \cite{keras}.

\section{Results and Discussion}

\subsection{Visual Classification of Wake Structures} \label{s:visualclass}

\begin{figure}[htbp]
\centering
	\begin{subfigure}[b]{0.49\textwidth}
	\centering
        \includegraphics[width=1\linewidth]{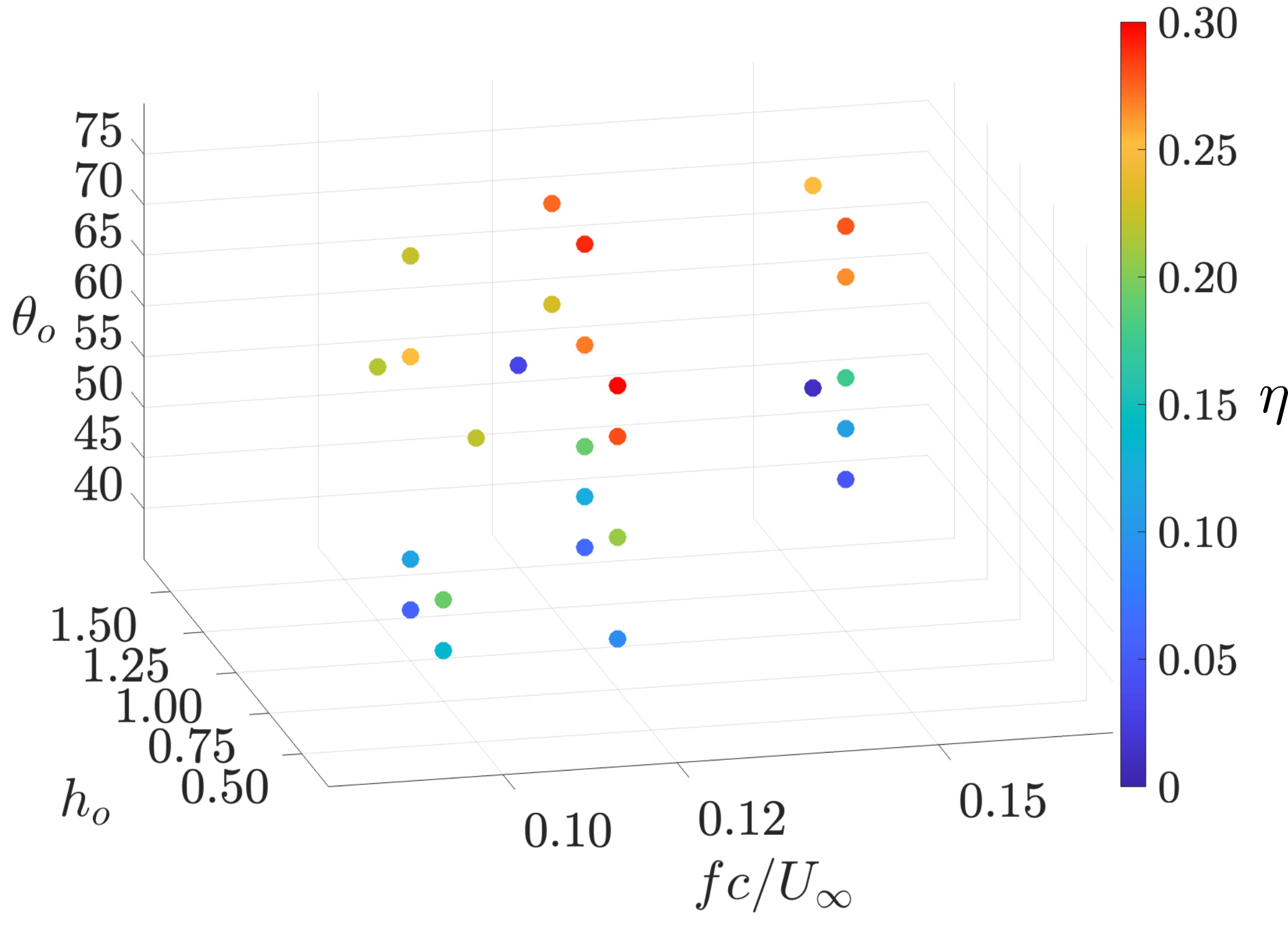}
        \caption{}
        \label{f:eta}
	\end{subfigure}
	\begin{subfigure}[b]{0.48\textwidth}
	\centering
        \includegraphics[width=1\textwidth]{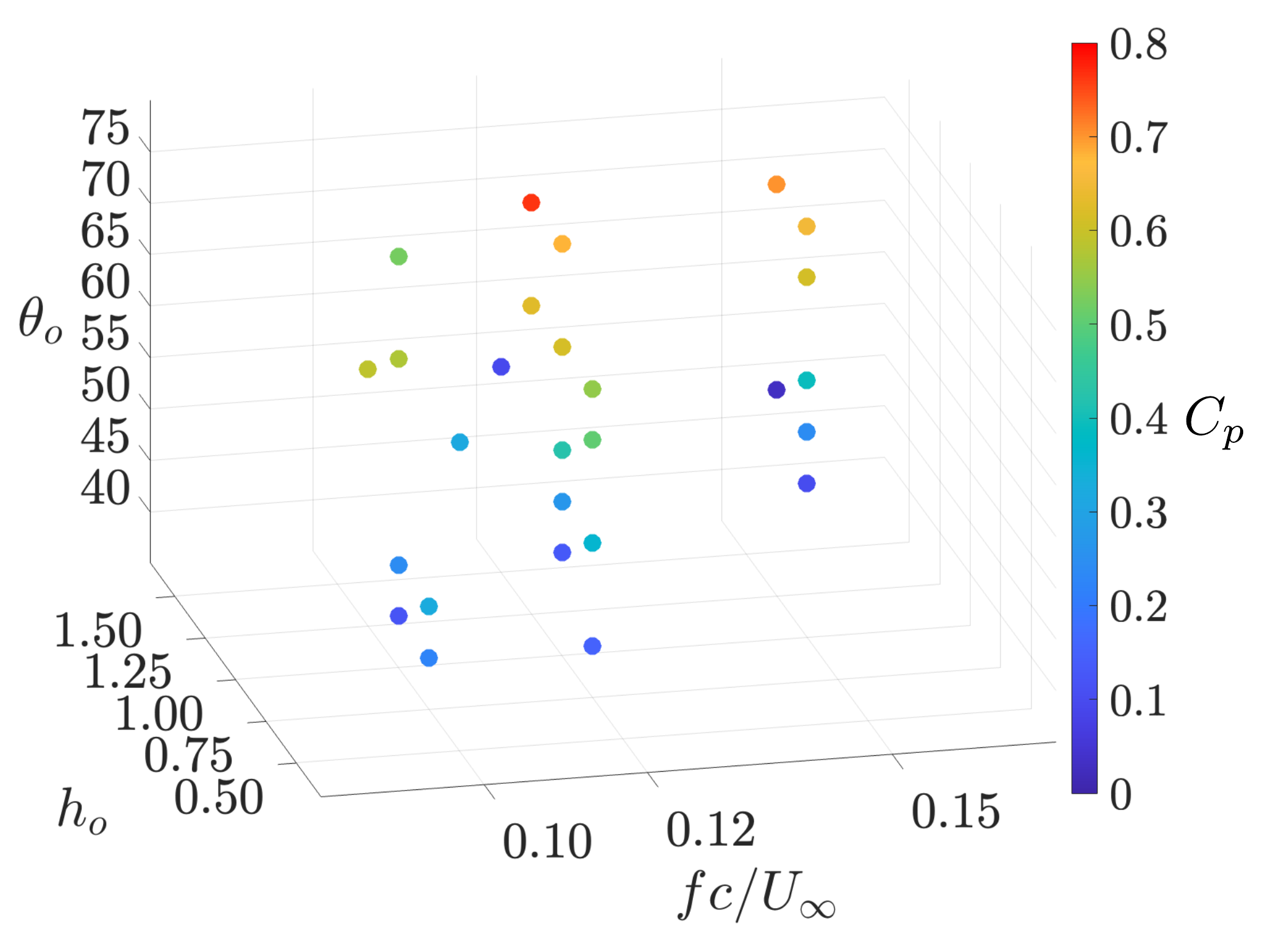}
        \caption{}
        \label{f:cp}
	\end{subfigure}
\caption{(a) Foil efficiency, $\eta$, and (b) Foil power coefficient, $C_p$, as a function of kinematic parameters.}
\label{f:etacp}	
\end{figure}

The kinematics explore a range of energy harvesting modes, whose optimal energy harvesting performance does not strongly correlate with a single pitch/heave amplitude or reduced frequency. Figure \ref{f:etacp} plots two performance metrics, efficiency and power coefficient, as a function of the three kinematic parameters. Depending on the metric there is a different range of optimal kinematics, and furthermore, some kinematics have approximately equal performance despite rather diverse parametric inputs. For efficiency, there is a trend for high pitch-heave amplitudes ($h_o=0.75-1.25; \theta_o=60^{\circ}-75^{\circ}$) with low-medium reduced frequency ($fc/U_\infty = 0.10; 0.12$) to perform better. However with the power coefficient metric, the heave amplitude becomes more important as the foil sweeps a greater available flow without increasing the denominator, resulting in optimal kinematics within the range $h_o=1.00-1.25; \theta_o=65^{\circ}-75^{\circ}$ and higher reduced frequencies ($fc/U_\infty = 0.12; 0.15$). This complexity leads to a range of realistic operating parameters for oscillating foils in energy harvesting, and motivates understanding the intricacies of the wake structure as it changes among these kinematics.

%Foil efficiency is also dependent on foil parameters. Figure \ref{f:etaparams} shows foil efficiency against the three foil parameters for all kinematics simulated. Overall, no clear efficiency trend can observed when comparing efficiency with respect to each foil parameter. The only conclusion from Figure \ref{f:etaparams} is that the most efficient cases are in a high pitch-heave amplitude range ($h_o=1.00-1.25; \theta_o=65^{\circ}-75^{\circ}$) and low reduced frequency ($fc/U_\infty = 0.10; 0.12$). However, few exceptions with high efficiency are observed the high reduced frequency of $fc/U_\infty=0.15$.

% \begin{figure}[htbp]
% \centering
% \includegraphics[width=0.5\textwidth]{figures/efficiency_scatter_NEW2.pdf}
% \caption{Foil efficiency against each foil parameter in all kinematics}
% \label{f:etaparams}
% \end{figure}

\begin{figure}[htbp]
\centering
\includegraphics[width=1\textwidth]{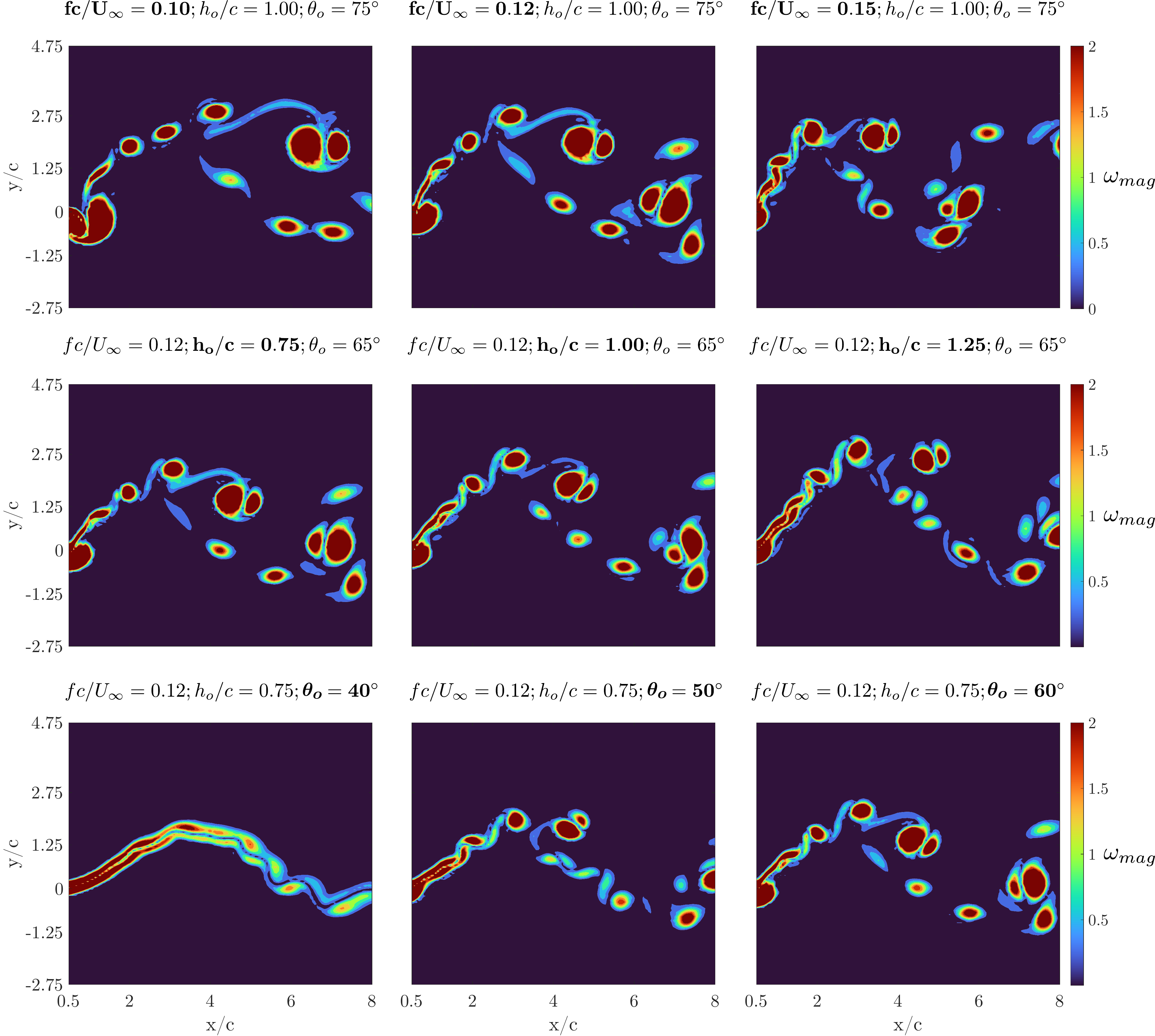}
\caption{Wake structures shown by vorticity magnitude ($\omega_{mag}$) for different combinations of kinematics. Top, middle and bottom rows show variation of reduced frequency, heave and pitch amplitudes, respectively, while the other parameters are held constant.}
\label{f:foileffects}
\end{figure}

Just as each kinematic parameter impacts performance, the vortex wake structure is also sensitive to changing kinematics. Figure \ref{f:foileffects} shows the vorticity magnitude within the wake of different foil kinematics. The top, middle and bottom rows vary the reduced frequency, heave and pitch amplitudes, respectively. With an increase in reduced frequency, the vortices shed from the foil remain closer to the foil when compared to lower frequencies (e.g. $fc/U_\infty=0.10$). This is due to the vortex formation time being larger than the half-stroke period as high reduced frequencies do not provide enough time for vortices to fully develop and shed \cite{RibeiroFranck2020}. When increasing the heave amplitude from $0.75$ and $1.25$, the vortices become closer to each other in the $y/c$ direction. Similar to increasing reduced frequency, the vortices are also smaller in size for high heave amplitude. The sequence of images in the bottom row with varying pitch amplitude perhaps provide the most articulated differences in the wake. For the pair of reduced frequency and heave amplitude illustrated, the low pitch amplitude of $40^{\circ}$ shows a shear layer and resulting instabilities, whereas for higher pitch amplitudes vortices are formed directly on the foil. 

\begin{figure}[htbp]
\centering
\includegraphics[width=0.45\textwidth]{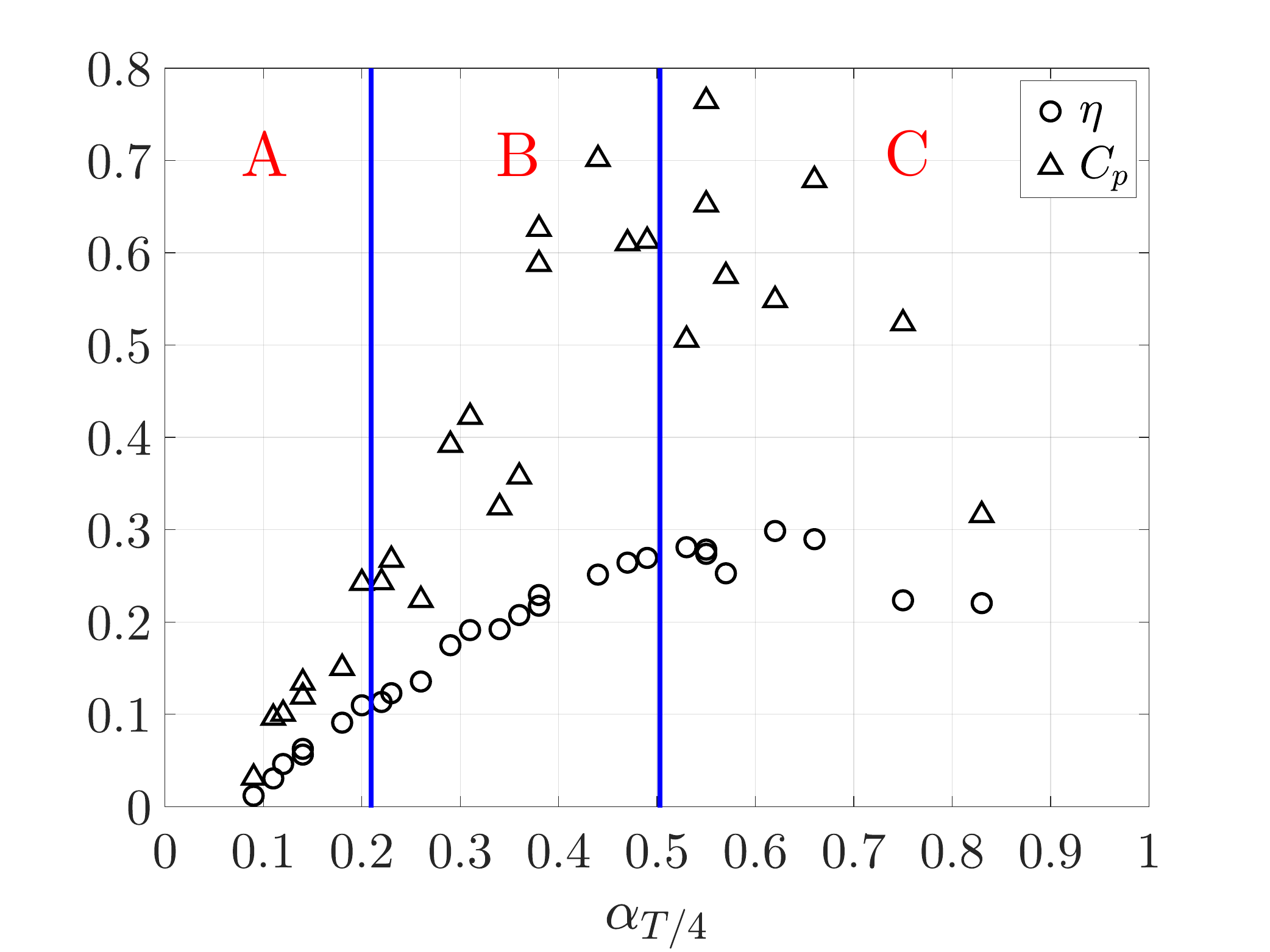}
\caption{Foil efficiency and power coefficient with respect to $\alpha_{T/4}$. Vertical lines indicate initial grouping of vortex wake structures.}
\label{f:etaalpha}
\end{figure}

As demonstrated in Figures \ref{f:etacp} and \ref{f:foileffects}, the large kinematic parameter space and the nonlinear correspondence with the performance metrics make it difficult to predict the vortex wake structure. Therefore, it is convenient to reduce the parameter space and combine reduced frequency, heave and pitch amplitudes into a single and representative parameter, namely the foil's relative angle of attack ($\alpha_{T/4}$), as defined by Equation \ref{eq:at4} \cite{RibeiroFranck2020}. Figure \ref{f:etaalpha} shows foil efficiency and power coefficient as a function of $\alpha_{T/4}$. There is a strong correlation between the foil efficiency and $\alpha_{T/4}$ as shown by the approximately linear trend from the $\alpha_{T/4}\approx0.09$ until $\alpha_{T/4}\approx0.50$. For kinematics surpassing $\alpha_{T/4}=0.50$, the foil efficiency shows more variation with respect to $\alpha_{T/4}$ and it starts to decrease after $\alpha_{T/4}\approx0.70$. Similarly, power coefficient also peaks at $\alpha_{T/4}=0.50$ but it shows larger variation as $\alpha_{T/4}$ is increased further.

%%%%% Without Cp
% \begin{figure}[htbp]
% \centering
% \includegraphics[width=0.45\textwidth]{figures/efficiency_vs_alpha.pdf}
% \caption{Foil efficiency with respect to all $\alpha_{T/4}$ values from all kinematics}
% \label{f:etaalpha}
% \end{figure}

Ribeiro et al. provide an initial attempt to divide the wake structures into self-similar groups by correlating the maximum vortex strength (measured by the Q criterion in a position immediately after it is formed on the suction side of the foil) with $\alpha_{T/4}$ \cite{RibeiroFranck2021}. These divisions, performed via visual inspection, are shown in Figure \ref{f:etaalpha}, and labeled `A', `B' and `C'. 
Although distinct differences are noted between these group divisions, the threshold criteria is solely predicted by the strength of one vortex and hence does not incorporate the positioning, strength and structure of all vortices. Furthermore, the criterion requires manual user input to determine when and where to probe for the vortex strength, and visual inspection in terms of where to draw the line divisions. Therefore, the proposed classification model proposed will verify how accurate the three visually labeled group divisions are drawn, and identify similarities in the wake structures not captured by the previous method. 

\subsection{Model Accuracy in Predicting Group Label of Test Data}

\begin{figure}[htbp]
\centering
        \includegraphics[width=0.72\linewidth]{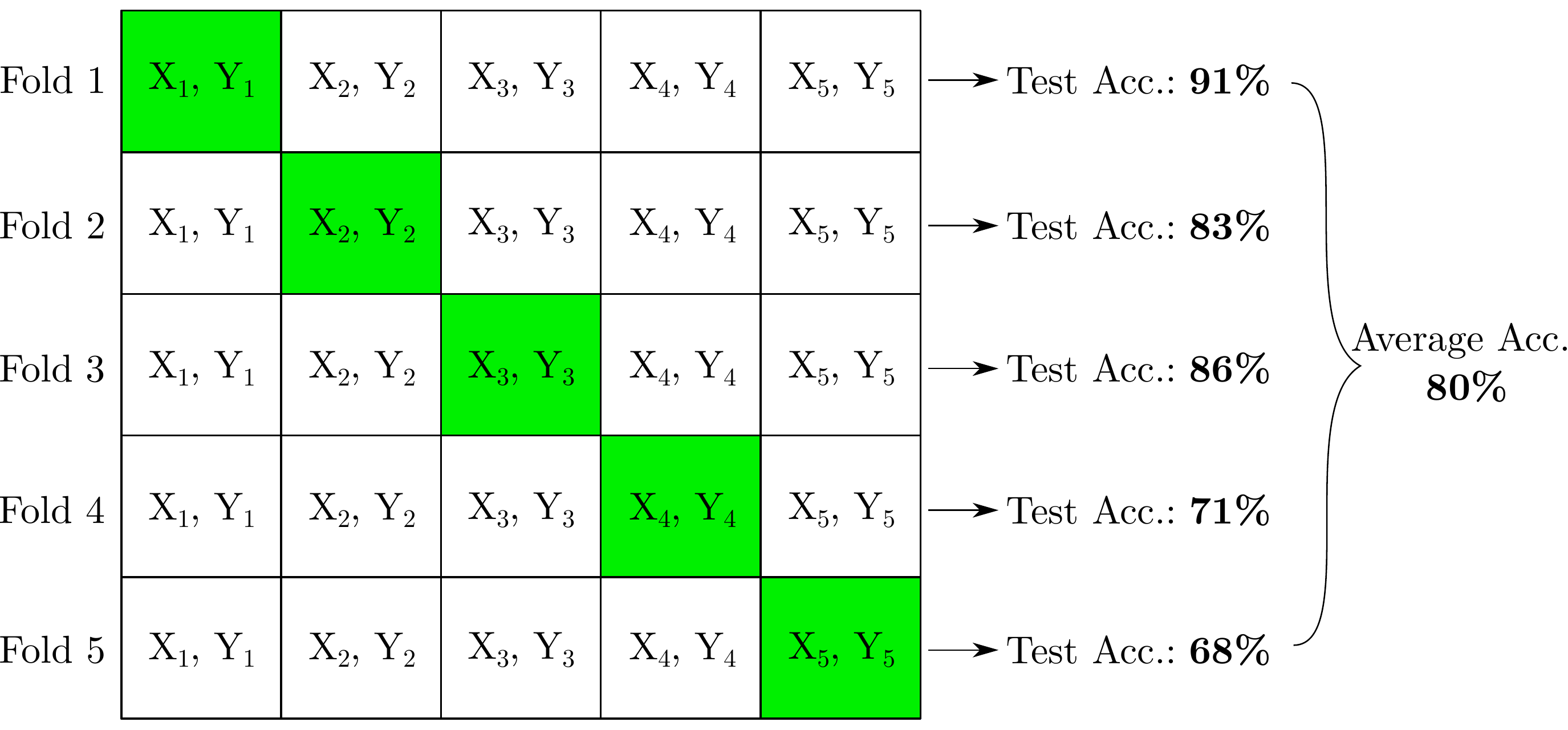}
\caption{Data setup through the 5-fold stratified cross-validation and test accuracy for each fold. The green squares represent the test data on each fold with $X$ standing for the matrices of vorticity magnitude values and $Y$, the visually assigned (actual) labels for each matrix. The other white squares are used for training.}
\label{f:datafold}	
\end{figure}

Using the $5$-fold stratified cross-validation as introduced in section \ref{s:modelarch}, Figure \ref{f:datafold} shows the test accuracy obtained for each fold and the combined accuracy among all folds. The test accuracy ranges from $68\%$ to $91\%$ and has an average accuracy of $80\%$. This percentage represents that $80\%$ of all samples processed by the algorithm are labelled the same as their visually inspected label.

\begin{figure}[htbp]
\centering
	\begin{subfigure}[b]{0.39\textwidth}
	\centering
        \includegraphics[width=0.8\linewidth]{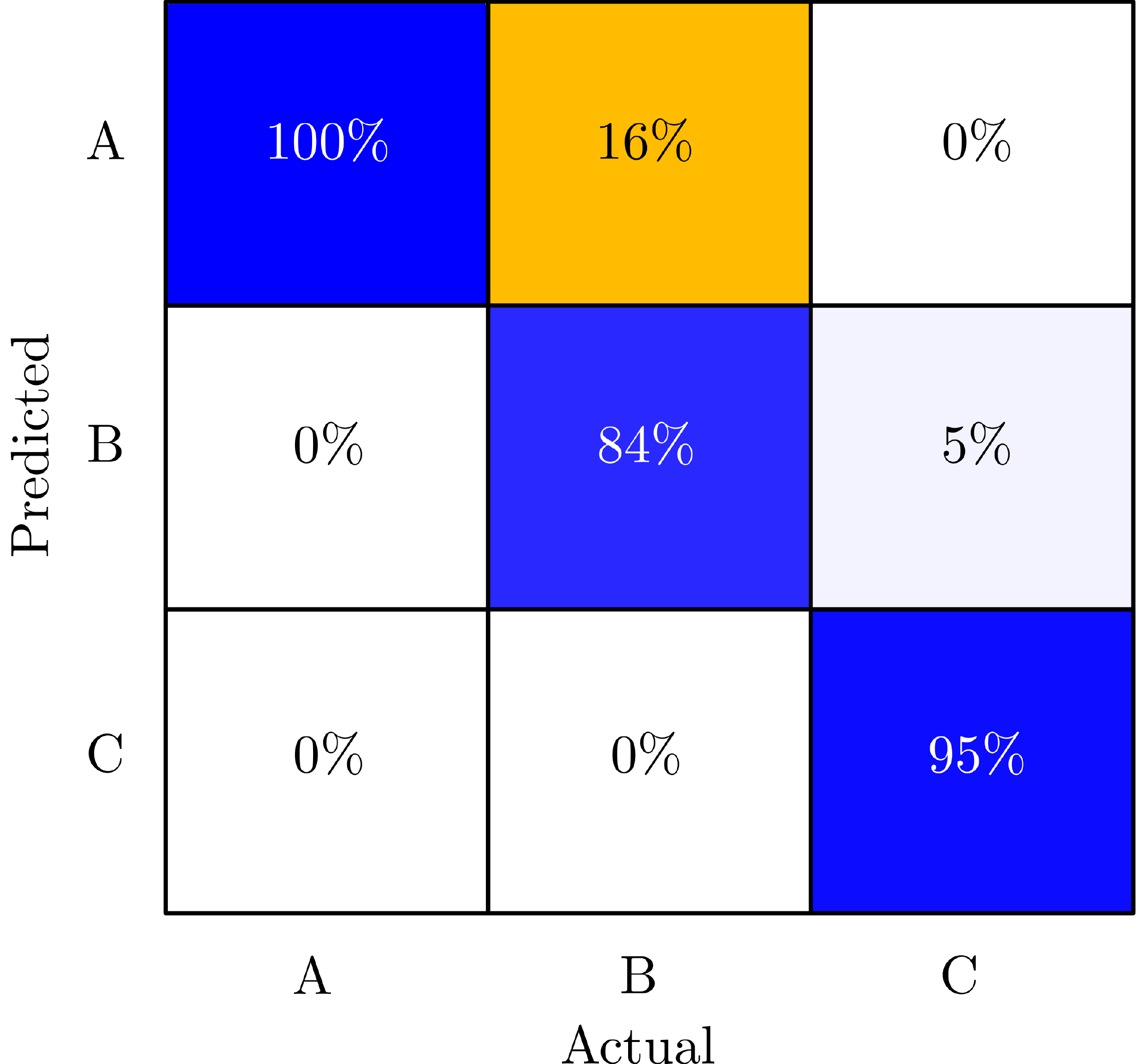}
        \caption{}
        \label{f:best}
	\end{subfigure}
	\begin{subfigure}[b]{0.60\textwidth}
	\centering
        \includegraphics[width=0.5\textwidth]{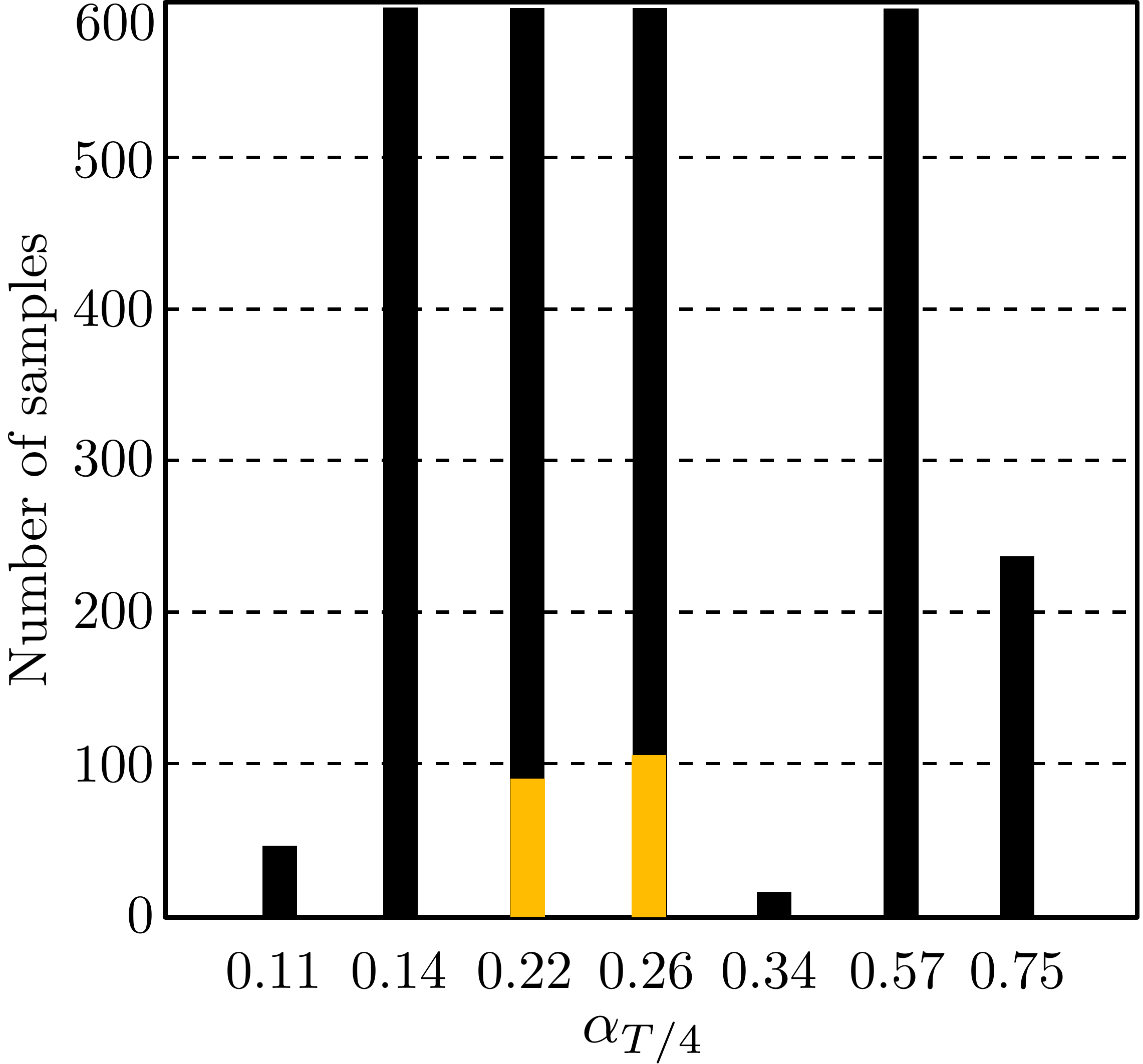}
        \caption{}
        \label{f:bestdist}
	\end{subfigure}
\caption{(a) Confusion matrix for the best performance (Fold 1) with mislabeled `A' samples colored in light orange (b) Test set distribution in number of samples among $\alpha_{T/4}$ values presented in fold 1 ($X_1$). The light orange bars represent the $16\%$ mislabeled data from the confusion matrix in (a).}
\label{f:bestanalysis}	
\end{figure}

To understand how each group performs in terms of predicting labels, a confusion matrix and the corresponding test data distribution for the best performance fold (fold 1) is shown in Figure \ref{f:bestanalysis}. The confusion matrix states the percentage values of how many visually labeled (actual) samples are predicted by the algorithm. Observing these values, the algorithm does not have any issues on discerning the kinematics within the group `A' and relatively small discrepancies are found for the groups `B' and `C', with only $5\%$ mislabeled samples in `C'. While the stratified cross-validation ensures a commensurate number of samples of each group in the training and test subsets, it does not ensure that each individual kinematics is represented in the test subset as shown by the test data distribution in Figure \ref{f:bestdist}. The light orange bars corresponds to the $16\%$ mislabeled samples that were classified as group `A' rather than `B'. Even though this is a relatively small percentage, the mismatched labels occurred in $\alpha_{T/4}$ values close to the boundary between these groups ($\alpha_{T/4}=0.20$).

\begin{figure}[htbp]
\centering
	\begin{subfigure}[b]{0.39\textwidth}
	\centering
        \includegraphics[width=0.8\linewidth]{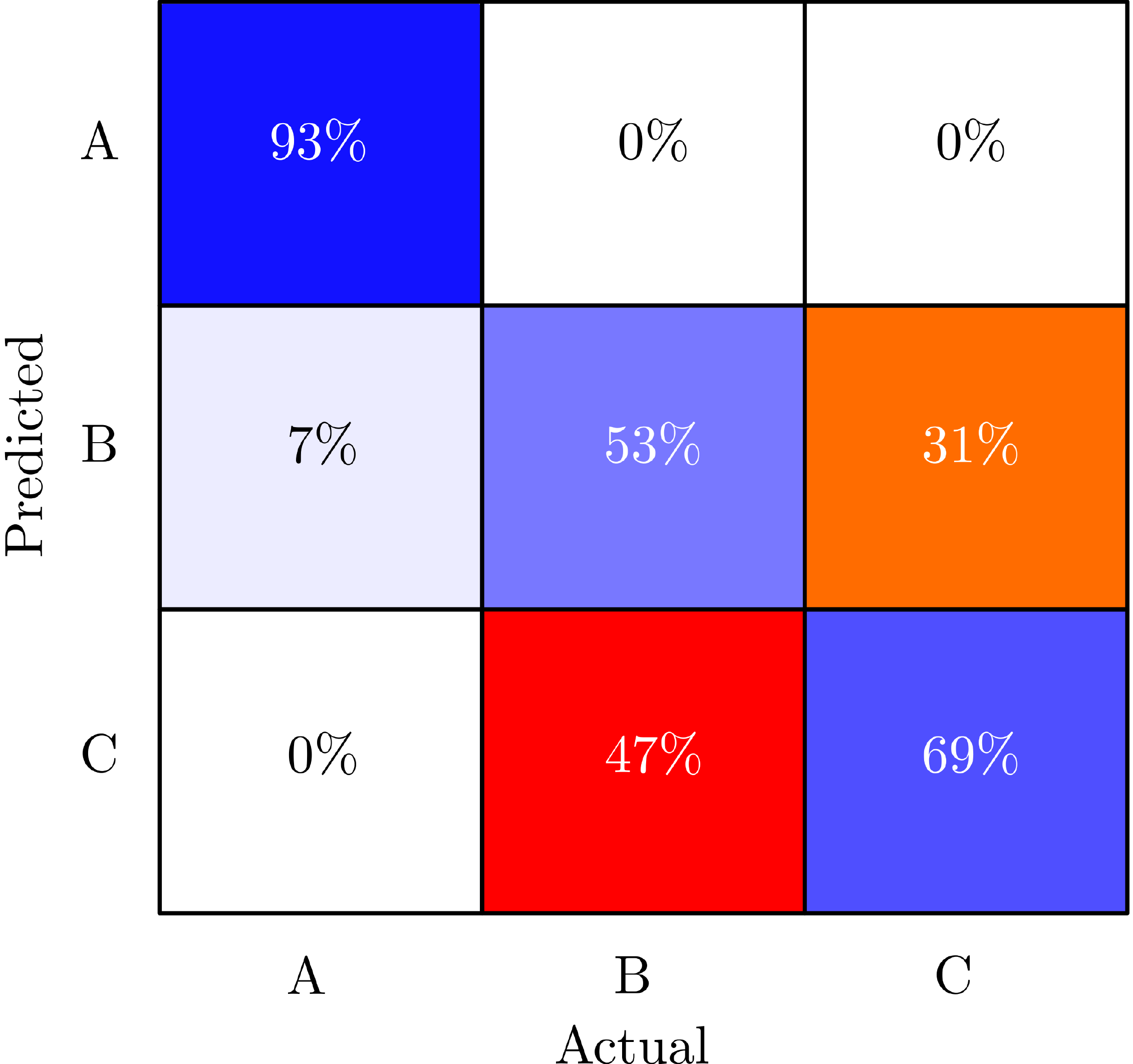}
        \caption{}
        \label{f:worst}
	\end{subfigure}
	\begin{subfigure}[b]{0.60\textwidth}
	\centering
        \includegraphics[width=0.5\textwidth]{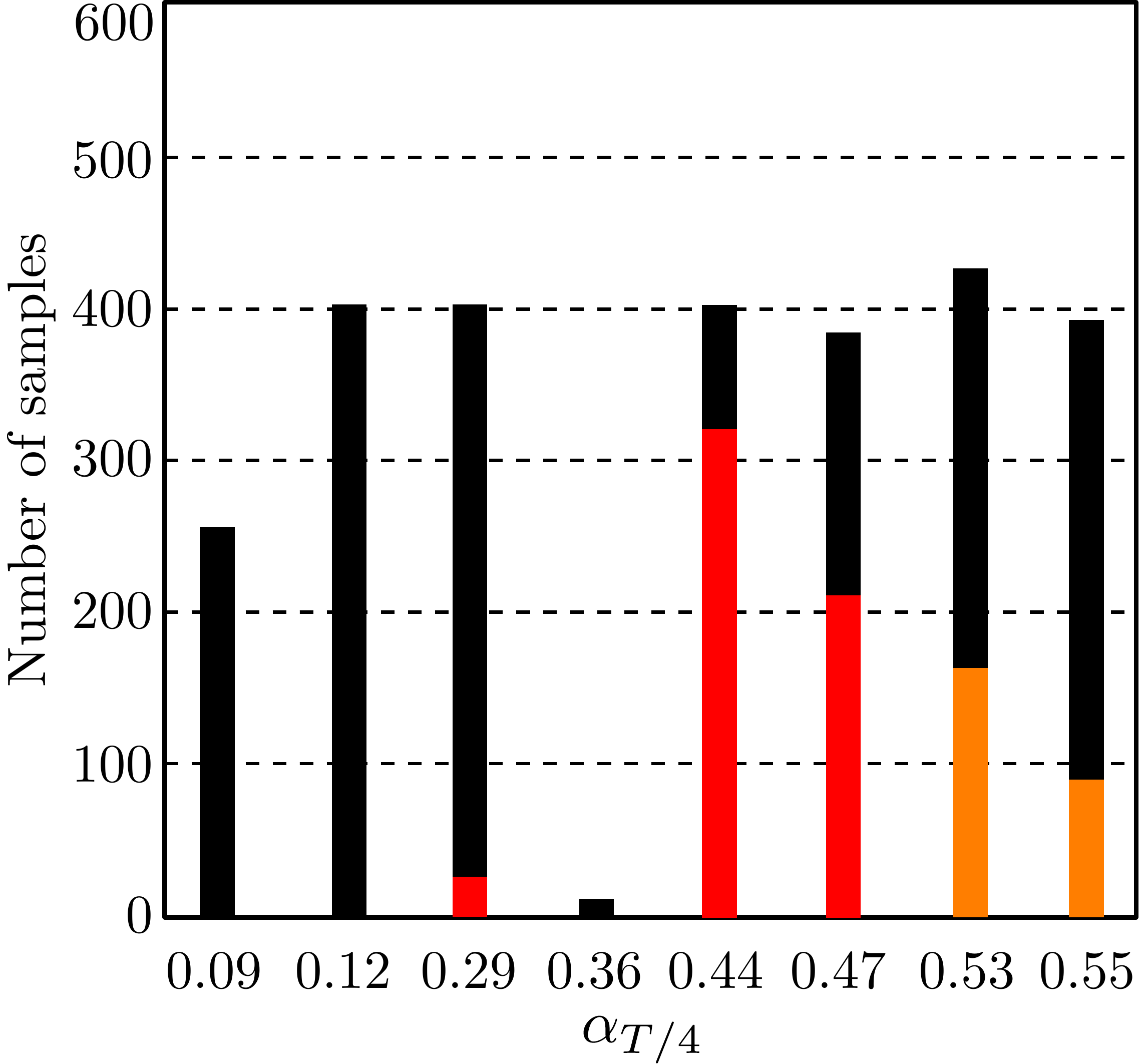}
        \caption{}
        \label{f:worstdist}
	\end{subfigure}
\caption{(a) Confusion matrix for the worst performance (Fold 5) with mislabeled `C' group colored in red and mislabeled `B' group in orange (b) Test set distribution in number of samples among $\alpha_{T/4}$ values presented in fold 5 ($X_5$). The red and orange bars represent the $47\%$ and $31\%$ mislabeled data from the confusion matrix in (a), respectively.}
\label{f:worstanalysis}	
\end{figure}

The fold with the worst performance (fold 5) articulates and continues to explore why some samples have different labels between the algorithm and what was visually inspected. Figure \ref{f:worst} shows the confusion matrix for fold 5. Even though this fold has the lowest test accuracy, group `A' is still being easily discerned by the algorithm. The consistent label prediction through the best and worst fold performances in group `A' may be due to the wake structure of shear layer vortices common for lower $\alpha_{T/4}$ values \cite{RibeiroFranck2021}. Between groups `B' and `C', the visually inspected labels and predicted labels start to deviate from one another. Although the majority of samples have the labels converging, the percentage of mismatched labels is significantly large, especially as group `B' samples being labelled as `C', which reaches $47\%$.

Similarly to what is shown for fold 1, Figure \ref{f:worstdist} shows the test data distribution among the kinematics tested in fold 5 and which kinematics correspond to the mismatched labels between that predicted by the algorithm and the visually inspected data. The red bars show that more than $80\%$ of the samples in $\alpha_{T/4}=0.44$ are classified as `C' and more than $50\%$ in $\alpha_{T/4}=0.47$ also have `C' as the predicted label. Regarding the other $\alpha_{T/4}$ values, the orange bars in Figure \ref{f:worstdist} show that the mislabeled $\alpha_{T/4}$ cases are close to the group boundary between `B' and `C' ($\alpha_{T/4}=0.50$), similar to the $\alpha_{T/4}=0.44, 0.47$ cases and to the mismatched labeled samples in fold 1, with the only difference being the confused groups are `A' and `B' for fold 1. This labeling confusion mainly found between `B' and `C' for fold 5 can be justified by the kinematics being closer to the visually inspected threshold between groups shown in Figure \ref{f:etaalpha} as the algorithm is discovering a strong wake similarity among these four cases ($\alpha_{T/4}=0.44; 0.47; 0.53; 0.55$).

Figures \ref{f:datafold}, \ref{f:bestanalysis} and \ref{f:worstanalysis} also show the capacity of the convolutional layers combined with LSTM units in the classification model on providing information of the wake to the user for various foil kinematics in the energy harvesting regime. The complex wake patterns in this regime are either discerned by the algorithm or each fold analysis gives a physical reason why the label mismatching occurred. Since most of the confused samples are located close to the group boundary between `B' and `C', an update is proposed on the group divisions.

%Although wakes may look similar behind oscillating foils in energy harvesting mode, the machine learning model presented in this paper either obtain high accuracy in discerning among the three proposed group labels or give a physical reason for why label mismatching happened on the worst performance fold. The average accuracy of $80\%$ is comparable with other classification models developed in literature on oscillating foils. For instance, Colvert et al. \cite{colvert2018} obtained an average accuracy of $90\%$ when discerning three wake types (2\textit{S}, 2\textit{P}+2\textit{S}, 2\textit{P}+4\textit{S}) behind oscillating foils in propulsion mode. Pollard et al. \cite{pollard2021} obtained a accuracy range of $70.6\% - 100\%$ when discerning 16 different wakes according to different Strouhal numbers in their oscillating foil system. It is important to emphasize the machine learning approaches were significantly different than the one developed in the current paper. Colvert et al. and Pollard et al. classified the wakes through the use of linear and nonlinear activation layers in a artificial neural network. The model described in the current study uses convolutional layers and takes advantage of the internal memories in LSTM units to try discerning the subtle differences in the wake behind oscillating foils in energy harvesting mode.

\subsection{Updating Group Boundaries Using Classification Model Results}

\begin{figure}[htbp]
\centering
\includegraphics[width=0.45\textwidth]{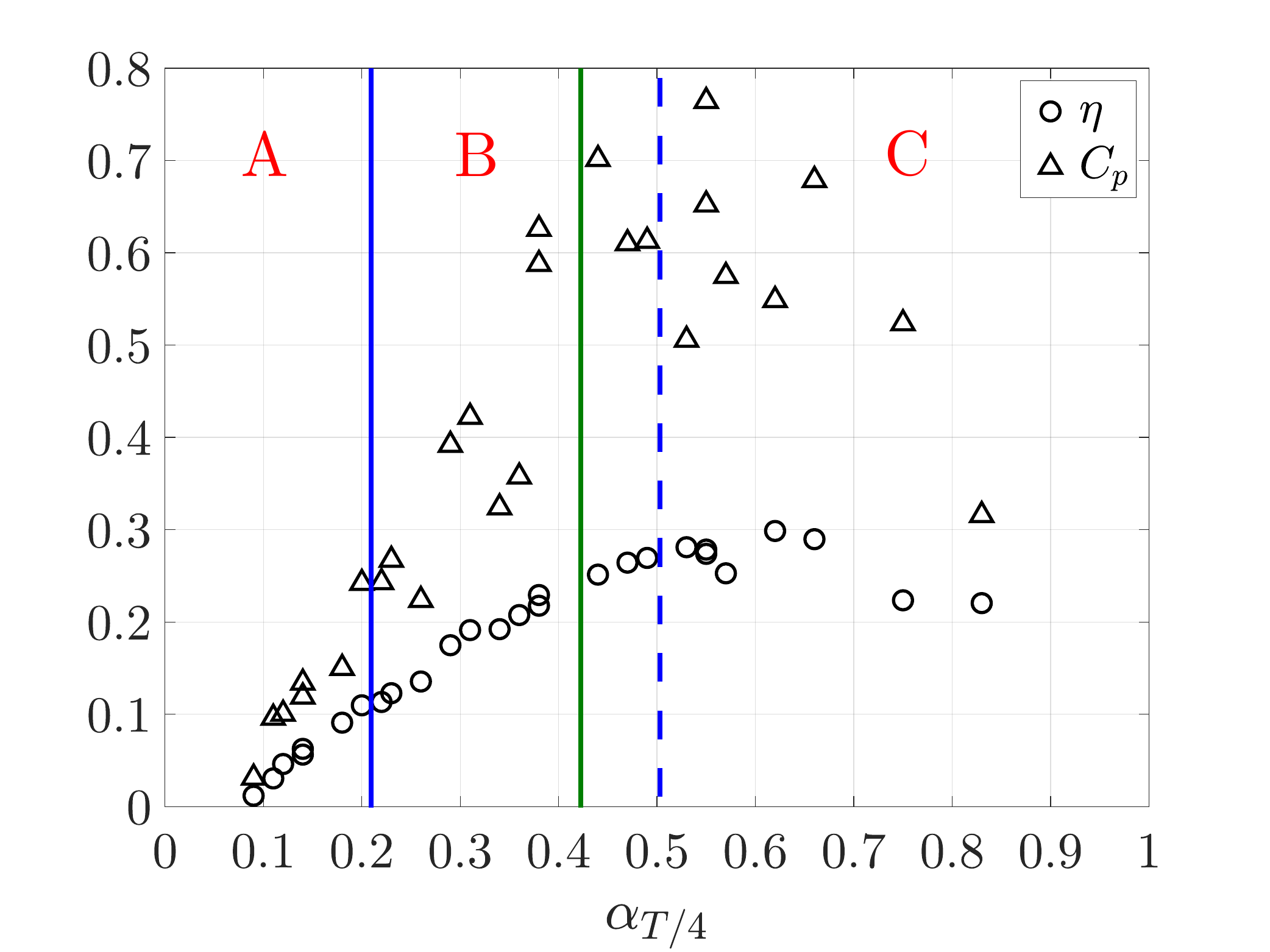}
\caption{Foil efficiency and power coefficient with respect to $\alpha_{T/4}$ based on Table \ref{t:kin} with the new group boundary between `B' and `C' (green). The dashed line is a reference of the original boundary location.}
\label{f:updatedetaalpha}
\end{figure}

To address the mismatch shown in the worst performance fold, Figure \ref{f:updatedetaalpha} shows the updated position of the group boundaries among `A', `B' and `C'. The updated boundary between `B' and `C' is now placed after $\alpha_{T/4}=0.44$ instead of $\alpha_{T/4}=0.50$ since the majority of samples in cases $\alpha_{T/4}=0.44, 0.47$ are labeled as `C' according to the algorithm (see Figure \ref{f:worstdist}).

%%%% Without Cp
% \begin{figure}[htbp]
% \centering
% \includegraphics[width=0.45\textwidth]{figures/efficiency_vs_alpha_updated.pdf}
% \caption{Foil efficiency with respect to various $\alpha_{T/4}$ based on Table \ref{t:kin} with the new group boundary between `B' and `C' (green). The dashed line is a reference the original boundary location.}
% \label{f:updatedetaalpha}
% \end{figure}

Proceeding with the updated group divisions, Table \ref{t:classification} compares the test accuracy results for each fold between the original and updated groups obtained after rerunning the model.

\begin{table}[htbp]
\setlength{\tabcolsep}{8pt}
\centering
\begin{tabular}{ccccccc}
\hline
\textbf{Group divisions} & \multicolumn{5}{|c|}{\textbf{Test accuracy of each fold}} & \textbf{Average} \\ \hline
Original & \multicolumn{1}{|c}{$91\%$} & $83\%$ & $86\%$ & $71\%$ & \multicolumn{1}{c|}{$68\%$} & $80\%$ \\
Updated & \multicolumn{1}{|c}{$91\%$} & $95\%$ & $98\%$ & $84\%$ & \multicolumn{1}{c|}{$84\%$} & $90\%$ \\
\hline \\
\end{tabular}
\caption{Test accuracy of each fold using original and updated group divisions.}
\label{t:classification}
\end{table}

As noticed in Table \ref{t:classification}, the updated group divisions provide higher accuracy compared to the original, increasing from $80\%$ to $90\%$ on average. This means the classification model better discerns the wake modes among the three new groups that were formed from analyzing the worst performing fold in the original groups. The new lowest performance fold has an accuracy of $84\%$ with the new group divisions, which is considerably higher than the $68\%$ in the original group divisions. More interestingly, all new folds obtained or maintained their accuracy (fold 1) compared to the original, with fold 3 reaching $98\%$. This result shows that even a considerably small change in group divisions can greatly affects the outcome of the classification model. With the new group divisions, the new folds are not equal to the original folds as the stratified cross-validation reorganize the data in order to maintain representative information of all groups in each test subset.

The machine learning model presented in this paper either obtained high accuracy in discerning among the three proposed group labels or gave a physical reason for why label mismatching happened on the worst performance fold. The average accuracy of $90\%$ after the update on group division boundaries is comparable with other classification models developed in literature on oscillating foils. Although using different methodologies, Colvert et al. \cite{colvert2018} also obtained an average accuracy of $90\%$ when distinguishing among three wake types (2\textit{S}, 2\textit{P}+2\textit{S}, 2\textit{P}+4\textit{S}) behind oscillating foils in propulsion mode. Pollard et al. \cite{pollard2021} obtained a accuracy range of $70.6\% - 100\%$ when discerning 16 different wakes according to different Strouhal numbers in an oscillating foil system. An important caveat is that Colvert et al. and Pollard et al. classified the wakes using linear and nonlinear activation layers in a artificial neural network, whereas the model described in the current study uses convolutional layers and LSTM units. 

Other changes to the classification algorithm would be likely improve accuracy. The update to group boundaries performed by analyzing fold 5 can be repeated to further refine boundary positioning or to propose new groups, such as another division in group `C', the largest of the three divisions. Another method to improve accuracy may be to give the algorithm more information (e.g. reduced frequency) as an input \cite{morimoto2021}. Lastly, instead of classification, an unsupervised clustering algorithm combined with a convolutional autoencoder may provide a different organization of wake modes since no visually (pre)labeled groups are needed \cite{calvet2020}.

\section{Conclusion}

A machine learning model is developed to classify wake structures behind an oscillating foil in the energy harvesting regime of kinematics. The goal of the paper is to utilize the machine learning algorithm to sort and classify various vortex wake modes using only the vorticity fields downstream of the oscillating foil. This model will give insight on wake similarity among various foil kinematics and correlate the kinematics with associated wake modes, which is important to build predictive models of oscillating foil arrays for energy harvesting.

Data is obtained through simulations of oscillating foils at $27$ unique kinematics, and time-dependent vorticity flow fields are extracted at equal times across three simulation cycles to form a total of $13,650$ samples. Based from previous work by Ribeiro et al. \cite{RibeiroFranck2021}, three initial classes are defined based on values of the relative angle of attack, $\alpha_{T/4}$ for each set of kinematics. The classification model consists of four convolutional layers and $90$ LSTM units applied on multiple input sequences of five samples each. The number of filters in the convolutional layers is decreasing as the model gets deeper with the goal to extract the most pertinent spatial features of the wake and then each sample is inserted into the LSTM units. The model's output consist of three neurons corresponding to the groups `A', `B', `C'.

The classification model obtained a test accuracy ranging from $68\%-91\%$, with a minimum accuracy of $93\%$ on separating the kinematics in group `A'. However, observing the worst fold performance, only $53\%$ of the labels in group `B' are predicted the same as the visually assigned label. The test data distribution for this fold showed that the respective $\alpha_{T/4}$ values that were mislabeled were all close to the predetermined boundary between groups `B' and `C'. 

Thus, a new group boundary of kinematics is defined to better separate groups `B' and `C' with cases having a medium $\alpha_{T/4}$ range ($0.20<\alpha_{T/4}<0.44$) and high range ($\alpha_{T/4}>0.44$). After rerunning the model with the new group divisions, the average model accuracy increased from $80\%$ to $90\%$, with one fold obtaining a $98\%$ accuracy. This result not only gives an insight on the expected wake structure among the described groups but also shows that a small change in the group boundary (from $\alpha_{T/4}=0.50$ to $\alpha_{T/4}=0.44$) can greatly affects classification performance.

%In future work, one can apply a CAE combined with clustering algorithm \cite{calvet2020} or provide additional information of the setup to the neural network \cite{morimoto2021} such as reduced frequency or pitch amplitude values in order to help improving accuracy on discerning wake modes of oscillating foils for energy harvesting. 

\section{Acknowledgments}

This work is supported by the National Science Foundation under Grant No. 1921594/1921359.

\bibliography{energyharv_ml}

\begin{thebibliography}{34}
\newcommand{\enquote}[1]{``#1''}
\providecommand{\natexlab}[1]{#1}
\providecommand{\url}[1]{\texttt{#1}}
\providecommand{\urlprefix}{URL }
\expandafter\ifx\csname urlstyle\endcsname\relax
  \providecommand{\doi}[1]{doi:\discretionary{}{}{}#1}\else
  \providecommand{\doi}{doi:\discretionary{}{}{}\begingroup
  \urlstyle{rm}\Url}\fi

\bibitem[{J. et~al.(2014)J., Lai, and Platzer}]{Young2014}
J., Y., Lai, J. C.~S., and Platzer, M.~F., \enquote{A review of progress and
  challenges in flapping foil power generation,} \emph{Progress in Aerospace
  Sciences}, Vol.~67, 2014.
\newblock \doi{10.1016/j.paerosci.2013.11.001}.

\bibitem[{Xiao and Zhu(2014)}]{Xiao2014}
Xiao, Q., and Zhu, Q., \enquote{A review on flow energy harvesters based on
  flapping foils,} \emph{Journal of Fluids and Structures}, Vol.~46, 2014.
\newblock \doi{10.1016/j.jfluidstructs.2014.01.002}.

\bibitem[{Williamson and Roshko(1988)}]{williamson1988}
Williamson, C. H.~K., and Roshko, A., \enquote{Vortex formation in the wake of
  an oscillating cylinder,} \emph{Journal of Fluids and Structures}, Vol.~2,
  1988.
\newblock \doi{10.1016/S0889-9746(88)90058-8}.

\bibitem[{Schnipper et~al.(2009)Schnipper, Andersen, and Bohr}]{schnipper2009}
Schnipper, T., Andersen, A., and Bohr, T., \enquote{Vortex wakes of a flapping
  foil,} \emph{Journal of Fluid Mechanics}, Vol. 633, 2009.
\newblock \doi{10.1017/S0022112009007964}.

\bibitem[{Koochesfahani(2012)}]{koochesfahani2012}
Koochesfahani, M.~M., \enquote{Vortical patterns in the wake of an oscillating
  airfoil,} \emph{AIAA Journal}, Vol.~27, 2012.
\newblock \doi{10.2514/3.10246}.

\bibitem[{Lai and Platzer(1999)}]{lai1999}
Lai, J. C.~S., and Platzer, M.~F., \enquote{Jet {Characteristics} of a
  {Plunging} {Airfoil},} \emph{AIAA Journal}, Vol.~37, 1999.
\newblock \doi{10.2514/2.641}.

\bibitem[{Ribeiro et~al.(2020)Ribeiro, Frank, and Franck}]{RibeiroFranck2020}
Ribeiro, B. L.~R., Frank, S.~L., and Franck, J.~A., \enquote{{Vortex dynamics
  and Reynolds number effects of an oscillating hydrofoil in energy harvesting
  mode},} \emph{Journal of Fluids and Structures}, Vol.~94, 2020.
\newblock \doi{10.1016/j.jfluidstructs.2020.102888}.

\bibitem[{Brunton et~al.(2020)Brunton, Noack, and Koumoutsakos}]{brunton2020}
Brunton, S.~L., Noack, B.~R., and Koumoutsakos, P., \enquote{Machine {Learning}
  for {Fluid} {Mechanics},} \emph{Annual Review of Fluid Mechanics}, Vol.~52,
  2020.
\newblock \doi{10.1146/annurev-fluid-010719-060214}.

\bibitem[{Lee and You(2019)}]{lee2019}
Lee, S., and You, D., \enquote{Data-driven prediction of unsteady flow over a
  circular cylinder using deep learning,} \emph{Journal of Fluid Mechanics},
  Vol. 879, 2019.
\newblock \doi{10.1017/jfm.2019.700}.

\bibitem[{Bhatnagar et~al.(2019)Bhatnagar, Afshar, Pan, Duraisamy, and
  Kaushik}]{bhatnagar2019}
Bhatnagar, S., Afshar, Y., Pan, S., Duraisamy, K., and Kaushik, S.,
  \enquote{Prediction of aerodynamic flow fields using convolutional neural
  networks,} \emph{Computational Mechanics}, Vol.~64, 2019.
\newblock \doi{10.1007/s00466-019-01740-0}.

\bibitem[{Hunt et~al.(1988)Hunt, Wray, and Moin}]{Hunt1988}
Hunt, J., Wray, A., and Moin, P., \enquote{Eddies, streams, and convergence
  zones in turbulent flows,} \emph{Center for Turbulence Research: Proceeding
  of the Summer Program}, 1988.
\newblock \urlprefix\url{https://ntrs.nasa.gov/citations/19890015184}.

\bibitem[{Jeong and Hussain(1995)}]{jeong1995}
Jeong, J., and Hussain, F., \enquote{On the identification of a vortex,}
  \emph{Journal of Fluid Mechanics}, Vol. 285, 1995.
\newblock \doi{10.1017/S0022112095000462}.

\bibitem[{Deng et~al.(2019)Deng, Wang, Liu, Wang, Li, and Liu}]{deng2019}
Deng, L., Wang, Y., Liu, Y., Wang, F., Li, S., and Liu, J., \enquote{A
  {CNN}-based vortex identification method,} \emph{Journal of Visualization},
  Vol.~22, 2019.
\newblock \doi{10.1007/s12650-018-0523-1}.

\bibitem[{Calvet et~al.(2021)Calvet, Dave, and Franck}]{calvet2020}
Calvet, A.~G., Dave, M., and Franck, J.~A., \enquote{Unsupervised clustering
  and performance prediction of vortex wakes from bio-inspired propulsors,}
  \emph{Bioinspiration \& Biomimetics}, Vol.~16, 2021.
\newblock \doi{10.1088/1748-3190/ac011f}.

\bibitem[{Kashir et~al.(2021)Kashir, Ragone, Ramasubramanian, Yurkiv, and
  Mashayek}]{kashir2021}
Kashir, B., Ragone, M., Ramasubramanian, A., Yurkiv, V., and Mashayek, F.,
  \enquote{Application of fully convolutional neural networks for feature
  extraction in fluid flow,} \emph{Journal of Visualization}, 2021.
\newblock \doi{10.1007/s12650-020-00732-0}.

\bibitem[{Morimoto et~al.(2021)Morimoto, Fukami, Zhang, Nair, and
  Fukagata}]{morimoto2021}
Morimoto, M., Fukami, K., Zhang, K., Nair, A.~G., and Fukagata, K.,
  \enquote{Convolutional neural networks for fluid flow analysis: toward
  effective metamodeling and low-dimensionalization,} \emph{arXiv preprint
  arXiv:2101.02535 [physics]}, 2021.

\bibitem[{Hochreiter and Schmidhuber(1997)}]{hochreiter1997}
Hochreiter, S., and Schmidhuber, J., \enquote{Long {Short}-{Term} {Memory},}
  \emph{Neural Computation}, Vol.~9, 1997.
\newblock \doi{10.1162/neco.1997.9.8.1735}.

\bibitem[{Han et~al.(2019)Han, Wang, Zhang, and Chen}]{han2019}
Han, R., Wang, Y., Zhang, Y., and Chen, G., \enquote{A novel spatial-temporal
  prediction method for unsteady wake flows based on hybrid deep neural
  network,} \emph{Physics of Fluids}, Vol.~31, 2019.
\newblock \doi{10.1063/1.5127247}.

\bibitem[{Nakamura et~al.(2021)Nakamura, Fukami, Hasegawa, Nabae, and
  Fukagata}]{nakamura2021}
Nakamura, T., Fukami, K., Hasegawa, K., Nabae, Y., and Fukagata, K.,
  \enquote{Convolutional neural network and long short-term memory based
  reduced order surrogate for minimal turbulent channel flow,} \emph{Physics of
  Fluids}, Vol.~33, 2021.
\newblock \doi{10.1063/5.0039845}.

\bibitem[{Zhang and Zhao(2020)}]{zhang2020}
Zhang, J., and Zhao, X., \enquote{A novel dynamic wind farm wake model based on
  deep learning,} \emph{Applied Energy}, Vol. 277, 2020.
\newblock \doi{10.1016/j.apenergy.2020.115552}.

\bibitem[{Colvert et~al.(2018)Colvert, Alsalman, and Kanso}]{colvert2018}
Colvert, B., Alsalman, M., and Kanso, E., \enquote{Classifying vortex wakes
  using neural networks,} \emph{Bioinspiration \& Biomimetics}, Vol.~13, 2018.
\newblock \doi{10.1088/1748-3190/aaa787}.

\bibitem[{Pollard and Tallapragada(2021)}]{pollard2021}
Pollard, B., and Tallapragada, P., \enquote{Learning hydrodynamic signatures
  through proprioceptive sensing by bioinspired swimmers,} \emph{Bioinspiration
  \& Biomimetics}, Vol.~16, 2021.
\newblock \doi{10.1088/1748-3190/abd044}.

\bibitem[{Li et~al.(2020)Li, Yang, Zhang, He, Deng, and Shen}]{li2020}
Li, B., Yang, Z., Zhang, X., He, G., Deng, B.-Q., and Shen, L., \enquote{Using
  machine learning to detect the turbulent region in flow past a circular
  cylinder,} \emph{Journal of Fluid Mechanics}, Vol. 905, 2020.
\newblock \doi{10.1017/jfm.2020.725}.

\bibitem[{Ribeiro et~al.(2021)Ribeiro, Su, Guillaumin, Breuer, and
  Franck}]{RibeiroFranck2021}
Ribeiro, B. L.~R., Su, Y., Guillaumin, Q., Breuer, K.~S., and Franck, J.~A.,
  \enquote{Wake-foil Interactions and Energy Harvesting Efficiency in Tandem
  Oscillating Foils,} \emph{Physics Review Fluids (accepted)}, 2021.
\newblock ArXiv:2103.05892.

\bibitem[{Kinsey and Dumas(2008)}]{kindum2008}
Kinsey, T., and Dumas, G., \enquote{Parametric Study of an Oscillating Airfoil
  in a Power-Extraction Regime,} \emph{AIAA Journal}, Vol.~46, No.~6, 2008, pp.
  1318--1330.
\newblock \doi{10.2514/1.26253}.

\bibitem[{Kim et~al.(2017)Kim, Strom, Mandre, and Breuer}]{Kim2017}
Kim, D., Strom, B., Mandre, S., and Breuer, K.~S., \enquote{Energy harvesting
  performance and flow structure of an oscillating hydrofoil with finite span,}
  \emph{Journal of Fluids and Structures}, Vol.~70, 2017, pp. 314--326.
\newblock \doi{10.1016/j.jfluidstructs.2017.02.004}.

\bibitem[{Issa(1986)}]{issa1986}
Issa, R.~I., \enquote{Solution of the implicitly discretised fluid flow
  equations by operator-splitting,} \emph{Journal of Computational Physics},
  Vol.~62, 1986.
\newblock \doi{10.1016/0021-9991(86)90099-9}.

\bibitem[{Weller et~al.(1998)Weller, Tabor, Jasak, and Fureby}]{weller1998}
Weller, H.~G., Tabor, G., Jasak, H., and Fureby, C., \enquote{A tensorial
  approach to computational continuum mechanics using object-oriented
  techniques,} \emph{Computers in Physics}, Vol.~12, 1998.
\newblock \doi{10.1063/1.168744}.

\bibitem[{Geuzaine and Remacle(2009)}]{gmsh}
Geuzaine, C., and Remacle, J.-F., \enquote{Gmsh: {A} 3-{D} finite element mesh
  generator with built-in pre- and post-processing facilities,}
  \emph{International Journal for Numerical Methods in Engineering}, Vol.~79,
  2009.
\newblock \doi{10.1002/nme.2579}.

\bibitem[{Abadi et~al.(2016)Abadi, Agarwal, Barham, Brevdo, Chen, Citro,
  Corrado, Davis, Dean, Devin, Ghemawat, Goodfellow, Harp, Irving, Isard, Jia,
  J{\'{o}}zefowicz, Kaiser, Kudlur, Levenberg, Man{\'{e}}, Monga, Moore,
  Murray, Olah, Schuster, Shlens, Steiner, Sutskever, Talwar, Tucker,
  Vanhoucke, Vasudevan, Vi{\'{e}}gas, Vinyals, Warden, Wattenberg, Wicke, Yu,
  and Zheng}]{tensorflow}
Abadi, M., Agarwal, A., Barham, P., Brevdo, E., Chen, Z., Citro, C., Corrado,
  G.~S., Davis, A., Dean, J., Devin, M., Ghemawat, S., Goodfellow, I.~J., Harp,
  A., Irving, G., Isard, M., Jia, Y., J{\'{o}}zefowicz, R., Kaiser, L., Kudlur,
  M., Levenberg, J., Man{\'{e}}, D., Monga, R., Moore, S., Murray, D.~G., Olah,
  C., Schuster, M., Shlens, J., Steiner, B., Sutskever, I., Talwar, K., Tucker,
  P.~A., Vanhoucke, V., Vasudevan, V., Vi{\'{e}}gas, F.~B., Vinyals, O.,
  Warden, P., Wattenberg, M., Wicke, M., Yu, Y., and Zheng, X.,
  \enquote{TensorFlow: Large-Scale Machine Learning on Heterogeneous
  Distributed Systems,} \emph{CoRR}, 2016.
\newblock ArXiv: abs/1603.04467.

\bibitem[{Chollet et~al.(2015)}]{keras}
Chollet, F., et~al., \enquote{Keras,} , 2015.
\newblock \urlprefix\url{https://github.com/fchollet/keras}, {P}ublisher:
  {GitHub}.

\bibitem[{Kingma and Ba(2015)}]{adam}
Kingma, D.~P., and Ba, J., \enquote{Adam: A Method for Stochastic
  Optimization,} \emph{3rd International Conference on Learning
  Representations. ICLR 2015 - Conf. Track Proc}, 2015.

\bibitem[{Fukami et~al.(2020)Fukami, Fukagata, and Taira}]{fukami2020}
Fukami, K., Fukagata, K., and Taira, K., \enquote{Assessment of supervised
  machine learning methods for fluid flows,} \emph{Theoretical and
  Computational Fluid Dynamics}, Vol.~34, 2020.
\newblock \doi{10.1007/s00162-020-00518-y}.

\bibitem[{Brunton and Kutz(2019)}]{bruntonbook}
Brunton, S.~L., and Kutz, J.~N., \emph{Data-Driven Science and Engineering:
  Machine Learning, Dynamical Systems, and Control}, Cambridge University
  Press, 2019.
\newblock \doi{10.1017/9781108380690}.

\end{thebibliography}

\end{document}